\documentclass[12pt]{article}
\usepackage{axodraw,graphicx,amsmath,amsthm,amscd,amsfonts}
\input xy
\xyoption{all}

\newtheorem{Proposition}[equation]{Proposition}

\theoremstyle{definition}

\newtheorem{Remark}[equation]{Remark}
\newtheorem{Example}[equation]{Example}

\newcommand{\Arr}{\mathbf{R}}
\newcommand{\Ell}{\mathbf{L}}

\newcommand{\Bord}[2]{(#1,#2)\text{--}\mathrm{Bord}}
\newcommand{\cE}{\mathcal{E}}
\newcommand{\cF}{\mathcal{F}}
\newcommand{\cG}{\mathcal{G}}

\renewcommand{\C}{\mathbb{C}}

\newcommand{\DC}[1]{D^{b} (#1)}
\newcommand{\dual}[1]{#1^{\vee}}

\newcommand{\coev}{\mathrm{coev}}
\newcommand{\ev}{\mathrm{ev}}

\DeclareMathOperator{\Ext}{Ext}

\DeclareMathOperator{\Hom}{Hom}
\renewcommand{\hom}{\Hom}
\newcommand{\heq}{\simeq}

\renewcommand{\i}{\infty}

\newcommand{\iso}{\cong}

\newcommand{\lotimes}{\otimes^{\Ell}}

\DeclareMathOperator{\Map}{Map}

\newcommand{\Rhom}{\Arr\mathrm{Hom}}
\newcommand{\RHom}{\Rhom}
\newcommand{\RGamma}{\Arr\Gamma}

\newcommand{\shfhom}{\mathit{Hom}}
\newcommand{\shfRhom}{\Arr\shfhom}

\newcommand{\Mod}[1]{\mathrm{Mod} (#1)}
\newcommand{\Ch}[1]{\mathrm{Ch} (#1)}

\renewcommand{\O}{\mathcal{O}}

\newcommand{\ptspace}{*}

\DeclareMathOperator{\Tor}{Tor}

\newcommand{\uln}[1]{\underline{#1}}

\newcommand{\xra}[1]{\xrightarrow{#1}}

\begin{document}

\hfill VPI-IPNAS 10-14

\vspace{1.0in}

\begin{center}

{\large\bf Two-dimensional topological field theories as taffy}

\vspace{0.2in}

Matt Ando$^1$, Eric Sharpe$^2$

\begin{tabular}{cc}
  \begin{tabular}{rl}
         $^1$ \hspace*{-5.5mm} & Department of Mathematics\\
                               & University of Illinois, Urbana-Champaign \\
                               & 1409 West Green Street \\
                               & Urbana, IL  61801\\
  \end{tabular} &
  \begin{tabular}{rl}
         $^2$ \hspace*{-5mm} & Physics Department\\
                             & Robeson Hall (0435) \\
                             & Virginia Tech\\
                             & Blacksburg, VA  24061\\
  \end{tabular}
\end{tabular}

{\tt mando@math.uiuc.edu}, {\tt ersharpe@vt.edu}

$\,$

\end{center}

In this paper we use trivial defects to define global taffy-like
operations on string worldsheets, which preserve the field theory.
We fold open and closed strings on a space $X$ into open strings on 
products of multiple copies of $X$,
and perform checks that the ``taffy-folded'' worldsheets have the
same massless spectra and other properties as the original
worldsheets.  Such folding tricks are a standard method in the
defects community; the novelty of this paper lies in deriving 
mathematical identities to check that {\it e.g.} massless spectra
are invariant in topological field theories.
We discuss the case of the B model extensively, and also derive
the same identities for string topology, where they become statements
of homotopy invariance.  We outline analogous results in
the A model, B-twisted Landau-Ginzburg models, and physical strings.
We also discuss the understanding of the closed string states 
as the Hochschild homology of the open string algebra, and outline
possible applications to elliptic genera.

\begin{flushleft}
October 2010
\end{flushleft}

\newpage

\tableofcontents

\newpage

\section{Introduction}

Defects in two-dimensional theories are forms of domain walls,
boundaries which connect open strings on potentially different spaces,
and so act as some type of two-dimensional domain wall,
see for example \cite{osh-affleck,bachasetal,roz1,roz2,khov-roz,br}.
Defects have appeared in numerous papers in the physics literature recently,
for a sample see for example
\cite{fiol,ksv,bdh,fss,dkr,s1,cr,kaps,kset} and references therein.  
They also seem to be implicit in parts of the mathematics literature,
in connection with ``enriched'' topological field theories 
described by higher categories (see {\it e.g.} \cite{lurietft}),
as we shall review.

In this paper we use `identity' (`trivial') defects to perform
taffy-like reparametrizations of string worldsheets, folding worldsheets
over themselves to create new, physically-equivalent but 
different-looking, worldsheet theories.

Using folding tricks locally around a defect is not new, indeed is
a standard method for computing spectra of operators in the
defects community.
What is (we believe) novel to this paper is the development of explicit
mathematical identities
required to carefully demonstrate that global foldings of worldsheets leave
the physics invariant.
For example,
given an open string on $X$, we develop mathematical identities required
to show that one gets physically-equivalent
open strings on products of any number of copies of $X$ by folding
and flattening along multiple identity defects.
One case of this is illustrated below:  
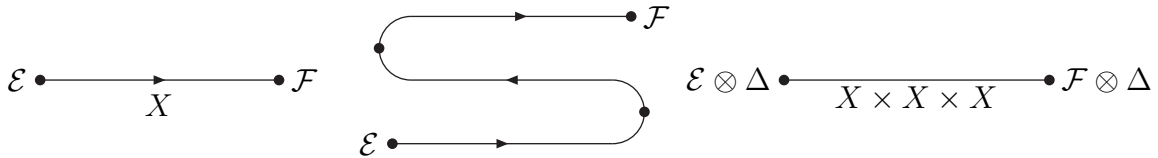
\begin{center}
\begin{picture}(120,100)
\ArrowLine(15,50)(105,50) 
\Vertex(15,50){2}  \Vertex(105,50){2}
\Text(60,45)[t]{$X$}
\Text(10,50)[r]{${\cal E}$}
\Text(110,50)[l]{${\cal F}$}
\end{picture}
$\: \:$
\begin{picture}(120,100)
\ArrowLine(15,26)(98,26)
\ArrowLine(98,50)(22,50)
\ArrowLine(22,74)(105,74)
\CArc(98,38)(12,-90,90)   \Vertex(110,38){2}
\CArc(22,62)(12,90,270)   \Vertex(10,62){2}
\Vertex(15,26){2}  \Vertex(105,74){2}
\Text(10,26)[r]{${\cal E}$}
\Text(110,74)[l]{${\cal F}$}
\end{picture}
$\: \:$
\begin{picture}(160,100)
\Line(30,50)(130,50)
\Vertex(30,50){2}  \Vertex(130,50){2}
\Text(80,48)[t]{$X \times X \times X$}
\Text(25,50)[r]{${\cal E}\otimes \Delta$}
\Text(135,50)[l]{${\cal F} \otimes \Delta$}
\end{picture}
\end{center}
The diagram above illustrates 
an open string on $X^3$ corresponding to an open string on $X$,
with boundaries determined in part by $\Delta$, a diagonal corresponding
to the identity defect.  From the general principles of folding,
one expects that these diagrams should describe equivalent physics.
We check that statement by comparing, for example, 
massless spectra in the B model,
deriving the identity
\begin{displaymath}
{\rm Ext}^*_X\left( {\cal E}, {\cal F} \right) \: = \:
{\rm Ext}^*_{X^3}\left( {\bf L} \pi_1^* {\cal E} \otimes^{\bf L}
\Delta_{23}^{\vee}, \Delta_{12} \otimes^{\bf L} {\bf L} \pi_3^* {\cal F} \right)
\end{displaymath}
confirming that the topological
field theory is unchanged, at the level of {\it e.g.}
D-branes described as objects in the derived category $D^b(X)$.
We can also fold closed strings along identity defects and flatten; 
for example,
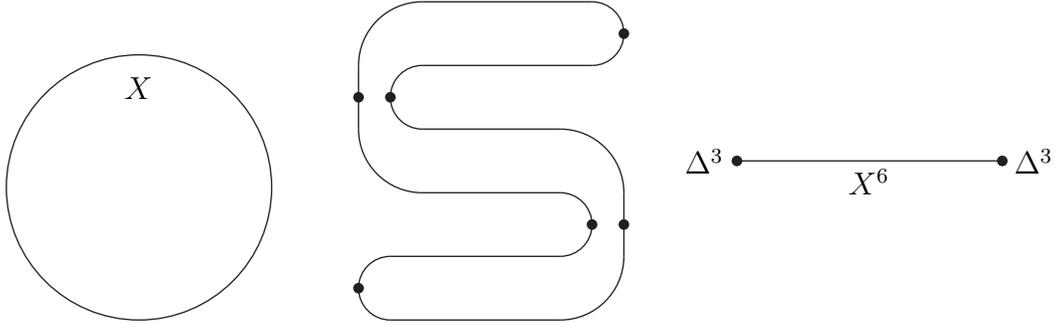
\begin{center}
\begin{picture}(120,120)(0,0)
\CArc(60,50)(50,0,360)
\Text(60,92)[t]{$X$}
\end{picture}
$\: \:$
\begin{picture}(120,120)   
\Line(22,0)(86,0)  \Line(22,24)(86,24)
\Line(34,48)(86,48)  \Line(34,72)(86,72)
\Line(34,96)(98,96)  \Line(34,120)(98,120)
\CArc(22,12)(12,90,270)   \Vertex(10,12){2}
\CArc(86,36)(12,-90,90)  \Vertex(98,36){2}
\CArc(86,48)(24,0,90)   \CArc(86,24)(24,-90,0)  \Vertex(110,36){2}
\Line(110,48)(110,24)
\CArc(98,108)(12,-90,90)  \Vertex(110,108){2}
\CArc(34,84)(12,90,270)   \Vertex(22,84){2}
\CArc(34,96)(24,90,180)   \CArc(34,72)(24,180,270)
\Line(10,72)(10,96)   \Vertex(10,84){2}
\end{picture} 
$\: \:$
\begin{picture}(140,120)
\Line(20,60)(120,60)
\Vertex(20,60){2}  \Vertex(120,60){2}
\Text(70,57)[t]{$X^6$}
\Text(15,60)[r]{$\Delta^3$}  \Text(125,60)[l]{$\Delta^3$}
\end{picture}
\end{center}
gives an open string on $X^6$ corresponding to a closed string on $X$.
In the case of the B model, we check the identity above by comparing
massless spectra:
\begin{displaymath}
H^*\left(X, \Lambda^* TX \right) \: = \:
{\rm Ext}^*_{X^6}\left( \Delta_{12}^{\vee} \otimes^{\bf L} \Delta_{36}^{\vee}
\otimes^{\bf L} \Delta_{45}^{\vee},
\Delta_{14} \otimes^{\bf L} \Delta_{23} \otimes^{\bf L} \Delta_{56} \right)
\end{displaymath}
checking that
topological field theories are, indeed, invariant under this
operation.
We can also include twists as we fold; for example,
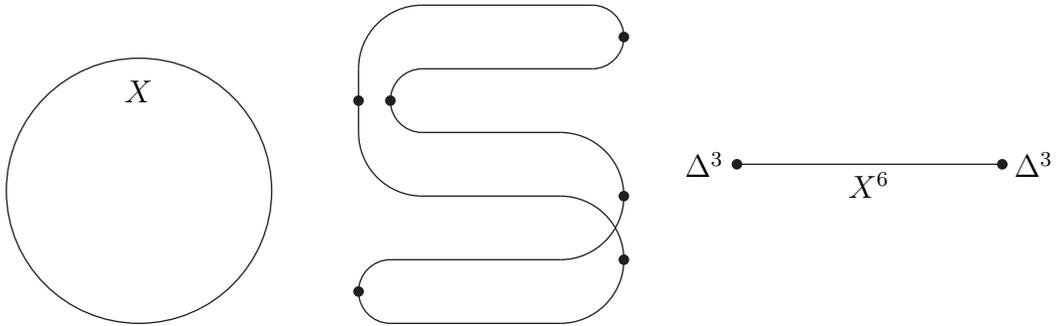
\begin{center}
\begin{picture}(120,120)(0,0)
\CArc(60,50)(50,0,360)
\Text(60,92)[t]{$X$}
\end{picture}
$\: \:$
\begin{picture}(120,120)
\Line(22,0)(86,0)  \Line(22,24)(86,24)
\Line(34,48)(86,48)  \Line(34,72)(86,72)
\Line(34,96)(98,96)  \Line(34,120)(98,120)
\CArc(22,12)(12,90,270)   \Vertex(10,12){2}
\CArc(86,48)(24,-90,90)   \Vertex(110,48){2}
\CArc(86,24)(24,-90,90)   \Vertex(110,24){2}
\CArc(98,108)(12,-90,90)  \Vertex(110,108){2}
\CArc(34,84)(12,90,270)   \Vertex(22,84){2}
\CArc(34,96)(24,90,180)   \CArc(34,72)(24,180,270)
\Line(10,72)(10,96)   \Vertex(10,84){2}
\end{picture}
$\: \:$
\begin{picture}(140,120)
\Line(20,60)(120,60)
\Vertex(20,60){2}  \Vertex(120,60){2}
\Text(70,57)[t]{$X^6$}
\Text(15,60)[r]{$\Delta^3$}  \Text(125,60)[l]{$\Delta^3$}
\end{picture}
\end{center}
gives a different open string on $X^6$ corresponding to the same closed
string on $X$.  In this case, demanding that B model states be invariant
under this operation implies the identity
\begin{displaymath}
H^*\left(X, \Lambda^* TX \right) \: = \:
{\rm Ext}^*_{X^6}\left( \Delta_{12}^{\vee} \otimes^{\bf L} \Delta_{36}^{\vee}
\otimes^{\bf L} \Delta_{45}^{\vee},
\Delta_{13} \otimes^{\bf L} \Delta_{24} \otimes^{\bf L} \Delta_{56} \right)
\end{displaymath}
which we check rigorously.

As these operations involve
folding worldsheets over onto themselves, we refer to them as
taffy operations.

For readers not acquainted with such folding tricks,
the philosophy is that inserting identity defects is, in principle,
a trivial operation, so these global foldings, these taffy operations we
describe, are no more than global reparametrizations, and so should encode
equivalent physics.  Much of this paper is devoted to carefully checking
that hypothesis.

One important special case of these constructions involves taking
a closed string on $X$ and constructing a physically-equivalent open
string on $X \times X$ by folding and flattening along two identity defects:
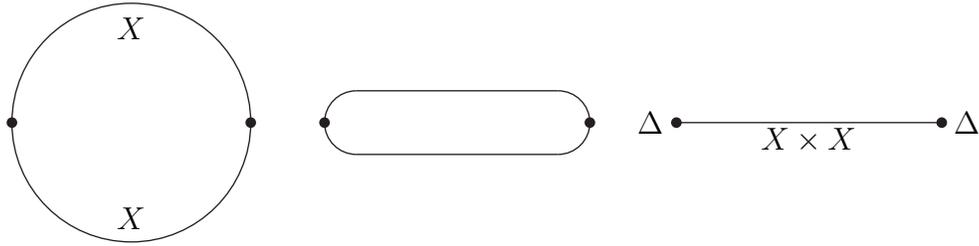
\begin{center}
\begin{picture}(100,100)(0,0)
\CArc(50,50)(45,0,180)
\CArc(50,50)(45,180,360)
\Vertex(95,50){2}
\Vertex(5,50){2}
\Text(50,10)[b]{$X$}
\Text(50,90)[t]{$X$}
\end{picture}
$\:  \:$
\begin{picture}(120,100)
\CArc(22,50)(12,90,270)
\CArc(98,50)(12,-90,90)
\Line(22,62)(98,62)   \Line(22,38)(98,38)
\Vertex(10,50){2}  \Vertex(110,50){2}
\end{picture}
$\:  \:$
\begin{picture}(120,100)
\Line(10,50)(110,50)
\Vertex(10,50){2}  \Vertex(110,50){2}
\Text(60,48)[t]{$X \times X$}
\Text(5,50)[r]{$\Delta$}
\Text(115,50)[l]{$\Delta$}
\end{picture}
\end{center}
(a manipulation well-known in the defects community).
The resulting identity relates Hochschild 
(co)homology (describing
closed string states) to endomorphisms of the open string algebra,
and for the B model 
is a well-known mathematical result, the ``Hochschild-Kostant-Rosenberg
isomorphism.''
Physically, this identity is a special case of the taffy
operations; mathematically, this
identity (and its analogues in other field theories)
is one of the foundations on which these taffy identities are
constructed.  We discuss this in detail, and in particular argue that
closed string states are more closely identified with
Hochschild homology, instead of Hochschild cohomology, of the open string
algebra.

The mathematical work in this paper for justifying
taffy operations amounts to using homological algebra identities to
reduce more complicated cases to the one above.
An important point is homotopy invariance, as explained in
section~\ref{hoch-closed}, which in some sense for this purpose
is more fundamental than
the Hochschild-Kostant-Rosenberg isomorphism or its analogues.
The homological algebra identities we derive are the expressions of
topological identities resulting from homotopy invariance.

Let us briefly illustrate what we mean by homotopy invariance.
Consider the $B$-model on a Calabi-Yau manifold $X$.  In that model,
$D$-branes correspond to elements $\cE$ of the derived
category $D^{b} (X)$ of bounded complexes of quasi-coherent
$\O_{X}$-modules.  Given two such complexes $\cE$ and
$\cF$, the massless states of open strings emanating from $\cE$ and
terminating on $\cF$ are measured by $\RHom (\cE,\cF)$, while closed
string states should be measured by the Hochschild cohomology
$\Hom_{X\times X} (\O_{X},\O_{X})$.     These two observations are
simply related as follows.
Consider a ``closeable configuration'' of open strings of the form 
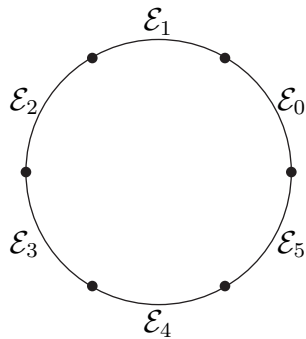
\begin{center}
\begin{picture}(120,120)
\CArc(60,60)(50,0,360)
\Vertex(110,60){2}
\Vertex(85,103){2}
\Vertex(35,103){2}
\Vertex(10,60){2}
\Vertex(35,17){2}
\Vertex(85,17){2}
\Text(105,87)[l]{${\cal E}_0$}
\Text(60,112)[b]{${\cal E}_1$}
\Text(15,87)[r]{${\cal E}_2$}
\Text(15,33)[r]{${\cal E}_3$}
\Text(60,8)[t]{${\cal E}_4$}
\Text(105,33)[l]{${\cal E}_5$}
\end{picture}
\end{center}
Given that the massless states for a string from $\cE$ to $\cF$ are
given by $\Rhom (\cE,\cF)$, it is natural to mathematically associate to this
configuration the group
\begin{equation}  \label{matt1}
    \Rhom (\cE_{0},\cE_{1})\times \Rhom (\cE_{1},\cE_{2})\times \dots 
\times \Rhom (\cE_{5},\cE_{0}).
\end{equation}
There is a certain redundancy in this description:  products of
open string states derived from 
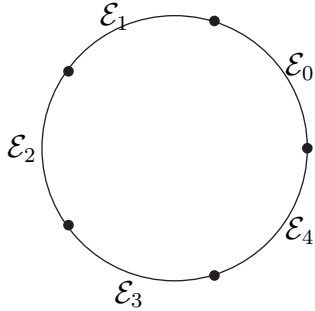
\begin{center}
\begin{picture}(120,120)
\CArc(60,60)(50,0,360)
\Vertex(110,60){2}
\Vertex(75,108){2}
\Vertex(20,89){2}
\Vertex(20,31){2}
\Vertex(75,12){2}
\Text(102,91)[l]{${\cal E}_0$}
\Text(43,110)[r]{${\cal E}_1$}
\Text(8,60)[r]{${\cal E}_2$}
\Text(43,10)[t]{${\cal E}_3$}
\Text(102,29)[l]{${\cal E}_4$}
\end{picture}
\end{center}
are related to those of the form~(\ref{matt1}) by the
composition\footnote{It is essential to use $\Rhom$ instead of $\Ext$
here.  As a related matter, composition of $\Rhom$ is defined only up to
homotopy, as we discuss in section \ref{prelims-r}.}
\[
    \Rhom (\cE_{4},\cE_{5})\times \Rhom (\cE_{5},\cE_{0}) \to \Rhom (\cE_{4},\cE_{0})
\]
in one direction, and by setting $\cE_{5} = \cE_{4}$ and inserting the
identity in the other.  There is a standard mathematical way for
encoding all this data, and, as we recall in
\S\ref{sec:string-topology}, Keller and McCarthy
\cite{McCarthy,Keller1,Keller2} 
show that one can extract the Hochschild  homology of $X$ from these data.  
This is part of what we mean by `homotopy invariance,' above.

This observation raises several questions.  For example, do diagrams
of the form above have any physical sense?  And can one use 
physical ideas to be more explicit about the relationship to
Hochschild homology?
We shall address these issues in this paper.

We begin in section~\ref{rev-defects} by briefly reviewing defects in
two-dimensional quantum field theories.
In section~\ref{bmodel:string-states} we work through the taffy operations
in detail for the B model, giving explicit descriptions and rigorously
checking identities for relating folded strings to the original strings.
Carefully studying these identities for the B model occupies the bulk of
this paper.
In section~\ref{string-top} we quickly rederive the same results in the
context of `string topology,' a mathematical abstraction of bosonic string
field theory.  In section~\ref{hoch-closed} we discuss old lore relating
closed string states to Hochschild homology (instead of cohomology)
of the open string algebra, and briefly outline some generalizations.
In sections~\ref{A-model}, \ref{matrix-fact},
\ref{critical-strings}, we
outline analogous results and conjectures in the A model, 
B-twisted Landau-Ginzburg models, and critical strings,
respectively.
We conclude in section~\ref{ell-gen} by outlining some conjectural applications
of this technology to elliptic genera.

\section{Review of defects}
\label{rev-defects}

\subsection{Physics basics}

A defect (see for example \cite{osh-affleck,bachasetal,roz1,roz2,khov-roz,br})
is an open string boundary that connects open strings on
two potentially distinct spaces.
Let $X$, $Y$ be two spaces, then a defect is an open-string-type boundary
defined by a submanifold of $X \times Y$, together with a bundle with
connection over that submanifold.
\begin{center}
\begin{picture}(120,20)
\Line(10,10)(110,10)
\Vertex(10,10){2}  \Vertex(110,10){2}   \Vertex(60,10){2}
\Text(35,7)[t]{$X$}    \Text(85,7)[t]{$Y$}
\end{picture}
\end{center}
Supersymmetry enforces the same conditions on that submanifold and bundle
as one would have in an ordinary open string -- for example, in the
B model topological field theory, one has a complex submanifold of $X \times Y$
with a holomorphic vector bundle.

Let us work through some examples.
Work locally on the complex plane, with a defect along the $x$-axis.
Consider for example a string with target ${\bf C}$ on each side of the defect.
We can describe this as a single complex boson $\phi$ defined over
the complex plane, or a pair $(\phi_U, \phi_L)$ on the upper-half-plane.
Boundary conditions such as, for example,
\begin{displaymath}
\lim_{y \rightarrow 0^+} \partial_n \phi \: = \: 0,
\: \: \:
\lim_{y \rightarrow 0^-} \partial_t \phi \: = \: 0
\end{displaymath}
or equivalently 
\begin{displaymath}
\left[ \begin{array}{cc}
\partial_n & 0 \\
0 & \partial_t \end{array} \right]
\left[ \begin{array}{c}
\phi_U \\ \phi_L \end{array} \right]
\: = \: 0
\end{displaymath}
which would be described more formally in the B model
by a sheaf on ${\bf C}^2$ with support
along $({\bf C}, 0) \subset {\bf C}^2$.
Boundary conditions such as
\begin{displaymath}
\lim_{y \rightarrow 0^+} \partial_n \phi \: = \:
\lambda \lim_{y \rightarrow 0^-} \partial_n \phi
\end{displaymath}
would be described more formally by a sheaf on ${\bf C}^2$ with
support along $\{ (\lambda x, x) | x \in {\bf C} \} \subset {\bf C}^2$.

Special cases of these boundary conditions can be described
by giving separate boundary conditions on $X$ and $Y$,
and a map between the corresponding branes, but the general case
is described by a sheaf on $X \times Y$.

One central idea from the study of defects, which we will use repeatedly
throughout this paper, is that of folding tricks (see for example
\cite{bachasetal} and references therein).
We have already used these implicitly in giving the data defining a defect.
Given open strings on $X$, $Y$, meeting along a defect, as
\begin{center}
\begin{picture}(120,20)
\Line(10,10)(110,10)
\Vertex(60,10){2}
\Text(35,7)[t]{$X$}    \Text(85,7)[t]{$Y$}
\end{picture}
\end{center}
we fold the combination of strings
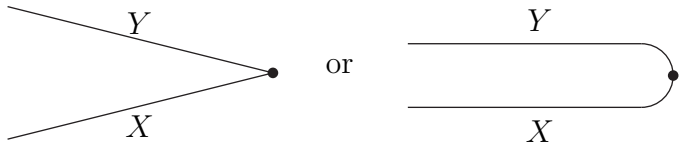
\begin{center}
\vspace{-35pt} \hfill \\
\begin{picture}(120,50)(0,25)
\Line(10,0)(110,25)  \Line(10,50)(110,25)
\Vertex(110,25){2}
\Text(60,9)[t]{$X$}  \Text(60,40)[b]{$Y$}
\end{picture}
$\:$ or $\:$
\begin{picture}(120,50)(0,25)
\Line(10,12)(98,12)
\CArc(98,24)(12,-90,90)  \Vertex(110,24){2}
\Line(98,36)(10,36)
\Text(60,7)[t]{$X$}  \Text(60,41)[b]{$Y$}
\end{picture}
\vspace{20pt} \hfill \\
\end{center}
into a single open string on the product $X \times Y$,
\begin{center}
\begin{picture}(120,20)
\Line(10,10)(110,10)
\Vertex(110,10){2}
\Text(60,7)[t]{$X \times Y$}
\end{picture}
\end{center}
so that the defect now appears as an ordinary boundary of an
ordinary open string on $X \times Y$.
This is one efficient way to understand why the data defining the
defect should be ordinary Chan-Paton factors on the product of the spaces,
and why the conditions for supersymmetry are the usual conditions, but
applied to Chan-Paton factors over the product of spaces.

There is one particular distinguished defect, the `identity' or
`trivial' defect, which joins open strings on $X$ to open strings also
on $X$.  The trivial defect is defined by trivial rank one 
Chan-Paton factors along
the diagonal embedding in $X \times X$.
It is so named because one can insert it into a string worldsheet without
changing the physics of that worldsheet theory.

In this paper, we insert identity defects and apply folding tricks globally
on worldsheets, not just locally in the neighborhood of a defect,
to give alternate (and physically equivalent) reparametrizations of
worldsheets.  We will check these methods extensively in the case of the
B model, at the level of boundaries defined by branes, antibranes, and
tachyons, as described mathematically by derived categories.
(In fact, not only will this allow us to work in significant generality,
but in addition, the formal manipulations required to check that
massless spectra are preserved become significantly easier when
working in derived categories than when working in, for example,
an ordinary category of coherent sheaves.)

It is unclear at present whether defects
can be consistently coupled to worldsheet gravity.
One potential problem is that for general defects, one often gets
infinite-dimensional moduli spaces of curves.  For example,
for a defect that starts in the  middle of a closed string worldsheet,
the shape of the defect is itself a modulus.
If one could consistently couple to worldsheet gravity, and define
critical string theories, the consequences appear to be somewhat radical.
For example, defects 
link open strings on distinct
spaces, and so, if (conformally invariant) defects
can be consistently coupled to worldsheet gravity,
and energy/momentum can flow across those defects, then those
defects would define nonlocal interactions in quantum gravity.
Just as a nonlocal interaction in ordinary quantum field theory violates
local energy/momentum conservation, this example of a nonlocal interaction
in quantum gravity would violate local energy/momentum conservation across
multiple spacetimes.
In any event, we shall not attempt to couple defects to worldsheet
gravity in this paper, and leave such considerations for other work.

\subsection{Typical mathematics application:  higher categories}
\label{higher-cat}

A typical mathematical application of defects is to give a physical
realization of descriptions of topological field theories using
higher categories, as in {\it e.g.} \cite{lurietft,costello}.
In this section we will briefly review such constructions and applications,
partly to contrast with the focus of this paper, which will be on taffy
operations with defects.

In the B model, a defect defines a Fourier-Mukai transform.
The data of the defect, an object ${\cal E}$ in $D^b(X \times Y)$,
defines the kernel of a Fourier-Mukai transform, hence a functor
\begin{displaymath}
F_{\cal E}( - ) \: \equiv \:
{\bf R} p_{Y *}\left( {\cal E} \otimes^{\bf L} {\bf L} p_X^* - \right)
\end{displaymath}
Such transforms can act on closed string states on either
space to generate closed string states on the other:
one pulls back a polyvector field
to the product, wedges with the Chern character of the object along the
defect, then pushes forward to the other space.
(This has been previously proposed in {\it e.g.} \cite{br}[pp 43-44].)

Given a defect on $X \times Y$ and another on $Y \times Z$,
if those two defect lines collide in parallel,
the result is a defect along $X \times Z$.
In the B model, which will be a focus of this paper,
this composition is believed to be defined
as follows \cite{andrei-utah}[prop. 5.1].
Consider $X \times Y \times Z$
with its projections $p_{12}$ to $X \times Y$,
$p_{13}$ to $X \times Z$, $p_{23}$ to $Y \times Z$.
Given ${\cal E}$ an object in $D^b(X \times Y)$ 
and ${\cal F}$ an object in $D^b(Y \times Z)$,
the composition is defined by the object
\begin{displaymath}
p_{13 *} \left( p_{12}^* {\cal E} \otimes p_{23}^* {\cal F} \right)
\end{displaymath}
in $D^b(X \times Z)$ (taking all operations to be derived).

In this language, there is an `identity' defect, defined by the diagonal
\cite{andrei-utah}[section 6.5].
If $\Delta_Y$ is the diagonal on $Y \times Y$, then for any
${\cal E}$ on $X \times Y$,
\begin{displaymath}
p_{13 *} \left( p_{12}^* {\cal E} \otimes
p_{23}^* \Delta_Y \right) \: \cong \: {\cal E}
\end{displaymath}

Suppose a defect line ends in the middle of a disk, as outlined below:
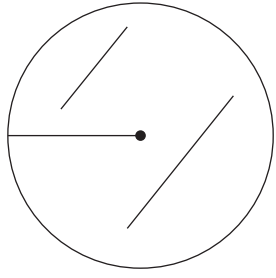
\begin{center}
\begin{picture}(100,100)
\CArc(50,50)(50,0,360)
\Line(0,50)(50,50)  
\Vertex(50,50){2}
\Line(45,15)(85,65)
\Line(20,60)(45,91)
\end{picture}
\end{center}
In the B model, if one is describing maps into a space $X$,
then along the defect line one has an object ${\cal E} \in
D^b(X \times X)$.  At the endpoint one naturally has a 
boundary-condition-changing operator defined by an element\footnote{
As a check, boundary-condition-changing operators inserted at the
intersection of the defect and the boundary would be counted,
in the B model, by elements
of ${\rm Ext}^*_{X^2}\left( \pi_1^* {\cal F} \otimes \pi_2^* {\cal F}^{\vee},
{\cal E} \right)$, where ${\cal F} \in D^b(X)$ defines the
disk boundary, and all operations assumed derived, conventions as
explained later in this text.  If we reel in the defect line, shrinking it
to zero length, then the states at the triple intersection are composed
with the states at the endpoint, namely elements of 
${\rm Ext}^*_{X^2}\left( {\cal E}, \Delta_* {\cal O}_X \right)$
(which for $X$ Calabi-Yau matches the groups above), 
to form elements of
\begin{displaymath}
{\rm Ext}^*_{X^2}\left( \pi_1^* {\cal F} \otimes \pi_2^* {\cal F}^{\vee},
\Delta_* {\cal O}_X \right) 
\: = \: {\rm Ext}^*_X\left( {\cal F}, {\cal F} \right)
\end{displaymath}
as expected for insertions along the disk boundary, 
using a mathematical identity we shall derive later in this paper.
} of
\begin{displaymath}
{\rm Ext}^*_{X \times X}\left( \Delta_* {\cal O}_X, {\cal E} \right)
\end{displaymath}
This has a natural categorical interpretation in terms of traces\footnote{
E.S. would like to thank T.~Pantev for explaining this material on
2-category traces.
} in
2-categories.  If we view $D^b(X)$ as a higher-categorical version of a
vector space (a 2-vector space), then $D^b(X \times X)$ is some version of
the space of endomorphisms of $D^b(X)$, and a trace map should be a map
to the analogue of the ground field, which in this case is 
$D^b({\bf C}-{\rm vect})$, the derived category of complex vector spaces.
In this language, the trace functor sends any object
${\cal E}$ in $D^b(X \times X)$ to 
\begin{displaymath}
{\rm Tr}\, {\cal E} \: \equiv \: {\bf R}{\rm Hom}_{X \times X}\left(
\Delta_* {\cal O}_X, {\cal E} \right)
\end{displaymath}
and any morphism $f: {\cal E} \rightarrow {\cal F}$ in $D^b(X \times X)$
is mapped to a morphism ${\rm Tr}(f)$ defined by
\begin{displaymath}
{\rm Tr}(f)(\phi) \: \equiv \: f \circ \phi
\end{displaymath}
This is a special case of the Ganter-Kapranov categorification of
the notion of trace to any 2-category \cite{ng}.

We should also note that open string defect diagrams seem to 
naturally correspond to ``string diagram'' pictures of 2-category
structures (see for example \cite{aw} for a discussion of the
mathematical notion of string diagrams).
Briefly, in what mathematicians call a string diagram, objects are represented
by 2-dimensional areas, 1-morphisms by boundaries between those areas,
and 2-morphisms by boxes or marked points along the boundaries.
Let us work through a simple example, both to illustrate string diagrams
and to display their correspondence to physics.
In the language of string diagrams, a commutative diagram
in 2-categories
\begin{displaymath}
\xymatrix{
X \ar@/^1pc/[rr]^{\cal F} \ar@/_1pc/[rr]_{\cal G} & \Downarrow \alpha  & Y
}
\end{displaymath}
would be represented by the (mathematical) string diagram
\begin{center}
\begin{picture}(100,100)
\Line(50,0)(50,45)   \Line(50,55)(50,100)
\Line(45,45)(45,55)  \Line(55,45)(55,55)
\Line(45,45)(55,45)   \Line(45,55)(55,55)
\Text(10,50)[l]{$X$}  \Text(90,50)[r]{$Y$}
\Text(52,25)[l]{${\cal G}$}   \Text(52,75)[l]{${\cal F}$}
\Text(50,48)[b]{$\alpha$}
\end{picture}
\end{center}
Physically, given open strings on $X$ and $Y$, defects joining the open
strings are defined (in the B model, say) by objects ${\cal F}, {\cal G}
 \in D^b(X \times Y)$,
each of which defines a Fourier-Mukai transform (hence a functor)
\begin{displaymath}
{\cal F}, {\cal G}: \: D^b(X) \: \longrightarrow \: D^b(Y)
\end{displaymath}
Furthermore, along the defect, one can insert an operator $\alpha$, 
which will be
an element of 
\begin{displaymath}
{\rm Ext}^*_{X \times Y}\left( {\cal F}, {\cal G} \right) \: = \:
H^* \left( {\bf R}{\rm Hom}_{X \times Y}\left( {\cal F}, {\cal G} \right)  
\right)
\end{displaymath}
and so will define a natural transformation between the two functors
defined by ${\cal F}$ and ${\cal G}$.  Clearly the mathematical `string diagram' above
corresponds very closely to the physical picture of corresponding open
strings joined along defects.

Although we will not work with mathematical 
string diagrams {\it per se} further in this paper, the picture above
lends itself to higher
categorical structures:  a defect along a $k$-codimensional
submanifold defines a $k$-morphism, and operators
on the defect and smaller defects define higher-order morphisms.
There is, in fact, a mathematical definition of (``enriched'') topological
field theories (see {\it e.g.} \cite{lurietft,costello,hopkins-lurie}), 
which involves representations
not just of the category of $d$-dimensional bordisms of
$d-1$-manifolds, but of appropriate $d$-categories $\Bord{0}{d}$
(actually,  ``$(\infty,d)$-categories'') of bordisms of manifolds of
dimension $0$ through $d$.

The Baez-Dolan cobordism hypothesis (now a family of theorems of Lurie 
\cite{lurietft}) says that an enriched topological field theory is
determined by its value on a point.  That is, to specify a
representation of $\Bord{0}{d}$ in a $d$-category $\mathcal{C}$ is
equivalent to specifying an object of  
$\mathcal{C}$, which is required to satisfy some hypotheses depending
on the precise flavor of enriched field theory one considers. 

For example, (\cite[Theorem 4.2.11]{lurietft}) to specify a
$0$---$2$-dimensional symmetric monoidal TFTs is equivalent to
specifying a ``Calabi-Yau'' object in a symmetric monoidal
$2$-category (actually symmetric monoidal ($\infty$,2)-category).
This is an object $X$ with a dual $\dual{X}$, an evaluation map 
\[
\ev: \:  X \otimes \dual{X}\xra{} \mathbf{1},
\]
a coevaluation map 
\[
\coev:  \: \mathbf{1}\xra{} X \otimes \dual{X}
\]
(satisfying various properties), and so a map, sometimes labelled
\[
   \dim (X) \: = \: \coev \circ \ev: \: \mathbf{1} \to \mathbf{1}
\]
This (strangely labelled) 
object $\dim (X)$ automatically carries an $S^{1}$-action
(because, despite the name,
it represents the massless states of a closed string).  Finally,
one should have an $S^{1}$-equivariant evaluation map 
\[
    \eta: \: \dim (X) \to \mathbf{1}.
\]
It is often equivalent (\cite[4.2.17]{lurietft}) to consider an
$S^{1}$-invariant cotrace $\mathbf{1} \to \dim (X).$

To obtain the B model on a Calabi-Yau $X$, consider the $2$-category
of differential graded categories, and for an object in it, take the
$(\infty,1)$--category of complexes of quasi-coherent $O_{X}$-modules on $X$ (whose
homotopy category is the derived category).  We interpret $\mathbf{1}$
as ${\cal O}_X$, the open string algebra, and the holomorphic top-form on
the Calabi-Yau defines a map $\mathbf{1} \rightarrow {\rm dim}\,X$.
For details, and an explanation of why $\infty$-categories are required,
see \S4.2 of   
\cite{lurietft}.

In this paper, we will be concerned with naively distinct mathematical
structures arising from defects, namely taffy operations.

\section{String states and taffy operations in the B model}

\label{bmodel:string-states}
 
In this section we will describe string states in strings with defects,
in the topological B model.
We will compute the spectrum of string states using folding tricks for
defects.  Folding tricks have been widely discussed in the defects
literature previously, see for example \cite{bachasetal} and references
therein.  Moreover, some of the mathematical consequences of
applying such folding tricks to topological field theories have
been previously described elsewhere, see for example \cite{hopkins-lurie}.
Our contribution is to work out and check explicit detailed
expressions relating various (folded) open and closed strings.

We should emphasize at this point that this `folding' is actually
a trivial operation on the string worldsheet -- in particular, despite
the name, no folds in the sense of non-differentiable structures are
being introduced into the maps into the target space.  Rather, this is
merely a reparametrization of the string worldsheet.

The result of this folding is to create 
physically-equivalent (but different-looking) string diagrams.
As a consistency check, we will demonstrate that massless spectra
of various physically-equivalent configurations agree.

\subsection{Operations on derived categories}   \label{prelims-r}

In our discussion of taffy-esque identities in the B model,
we will be working extensively with objects in derived categories.
When we perform operations such as duals, tensor products,
and homomorphisms, we necessarily mean derived duals, derived tensor
products, and so forth. 
As these concepts are not widely used in the physics community,
in this section we briefly recall some of the basic facts about
derived categories and the derived operations we use in this paper.  
We would also like to take this opportunity to thank A.~Caldararu
for numerous discussions of the identities in this and the next
subsection.

Let $X$ be a ringed space, such as a Calabi-Yau manifold, with
structure sheaf $\O_{X}$.  By ``$\O_{X}$-module'' we will
always mean a quasi-coherent sheaf of $\O_{X}$-modules.  We write
$\Mod{X}$ for the category of $\O_{X}$-modules.    If $E$, $F$,
and $G$ are $\O_{X}$-modules, then we can form the $\O_{X}$-modules
$E\otimes_{X} F$, 
$\shfhom_{X} (F,G)$, and $E^{*} = \shfhom_{X} (E,O_{X}).$   The sheaf
$\shfhom_{X} (F,G)$ is sometimes called the ``internal hom'', because
it is itself an $\O_{X}$-module: it is internal to the category of
$O_{X}$-modules.  Also, we can form the ring $R=\Gamma (X,\O_{X})$     and
the $R$-module $\Gamma E.$  The essential properties of these
operations are expressed by the identities 
\begin{equation}\label{eq:7}
\begin{split} 
     \shfhom_{X} (E\otimes_{X} F,G) \: & \iso \: \shfhom_{X} (E,\shfhom_{X}
     (F,G)) \\
     E^{*}\otimes_{X} F \: & \iso \: \shfhom_{X} (E,F) \text{ if }E\text{ is
     finitely generated}\\
     (E^{*})^{*} \: & = \: E  \text{ if }E\text{ is
     finitely generated}\\
     E^{*}\otimes F^{*} \: & = \: (E\otimes F)^{*}  \text{ if }E \text{ and } F
     \text{ are
     finitely generated}\\
     E \: & \iso \: \shfhom_{X} (\O_{X},E) \\
     \Hom_{\Mod{X}} (E,F) \: & \iso \: \Gamma (X,\shfhom (E,F)) \text{ and so } \\
     \Gamma (X,E) \: & \iso \: \Gamma (X,\shfhom (\O_{X},E)).
  \end{split}
\end{equation}

Let $f: X \to Y$ be a map of ringed spaces.  Associated to these we
have operations 
\[
\xymatrix{ 
           {\Mod{Y}} \ar@<1.5ex>[r]^{f^{*}} & 
           {\Mod{X}} 
                     \ar@<+1.5ex>[l]^{f_{*}}
}
\]
Some essential properties of these operations are\footnote{
The second line is a slight abuse of notation.
The point is that if
$\ptspace$ is a point, then $\O_{\ptspace}$ is the sheaf whose value
global sections are just the complex numbers.  Taking global sections
is an equivalence of categories between $\O_{\ptspace}$-modules and
$\C$-modules.  On the left one has an $\O_{C}$-module, on the right
one has the corresponding $\C$-module.
} 
\begin{align*}
     f_{*}E (U) \: & = \: E (f^{-1} (U)) \text{ if } U\text{ is an open set of
     }Y.  \text{ In particular}\\
     f_{*}E \: & \iso \: \Gamma (X,E) \text{ if } Y \text{ is a one-point space.}\\ 
     \Hom_{\Mod{X}} (f^{*}E,F) \: & \iso \: \Hom_{\Mod{Y}} (E,f_{*}F) \\
     f^{*} (E\otimes F) \: & = \: f^{*}E \otimes f^{*}F\\
     f^{*} (E^{*}) \: & = \: (f^{*}E)^{*}
\end{align*}

Let $\DC{X}$ be the the derived
category of bounded complexes of quasi-coherent $\O_{X}$-modules.   
An $\O_{X}$-module $E$ gives rise to an object of $\DC{X}$,
namely the complex $\dots 0 \to E \to 0 \dotsb $ which is $E$ in degree
$0$ and $0$ in other degrees.  We simply write $E$ for this object of
$\DC{X}.$    The morphisms in the derived category 
from $\cE$ to $\cF$ form a cochain complex, denoted $\Rhom_{X}
(\cE,\cF).$   The derived category has a tensor product, denoted $\cE
\lotimes_{X}\cF$, an internal Hom, denoted $\shfRhom_{X} (\cE,\cF)$, and a
dual $\dual{\cE}=\shfRhom (\cE,\O_{X}).$    Moreover, given a complex
of $\O_{X}$-modules $\cE$ we can form the chain complex of $\Gamma
(X,\O_{X})$-modules  $\RGamma (X,\cE)$.  These constructions are related by
identities much like those for $\Mod{X}$, namely
\begin{align*}
     \RHom_{X} (\cE\otimes_{X} \cF,\cG) \: & \heq \: \RHom_{X} (\cE,\shfRhom_{X}
     (\cF,\cG)) \\
     \cE^{\vee}\lotimes_{X} \cF \: & \heq \: \shfRhom_{X} (\cE,\cF)\text{ if
     }\cE\text{ is 
     finitely generated}\\
     \cE \: & \heq \: \shfRhom_{X} (\O_{X},\cE) \\
     \RHom_{X}(\cE,\cF) \: & \heq \: \RGamma \shfRhom (\cE,\cF) \\
     \RGamma \cE \: & \heq \: \RGamma \shfRhom (\O_{X},\cE).
\end{align*}

In the above, the symbol $\heq$ denotes quasi-isomorphism, meaning the
cochain complexes have isomorphic cohomology.   Familiar cohomological
invariants may be accessed via the identities
\begin{align*}
     \Ext^{i}_{X} (\cE,\cF) \: & = \: H^{i}\Rhom_{X} (\cE,\cF)  \\
     H^{i} (X;\cE) \: & = \: H^{i}\RGamma (\cE) \iso \Ext^{i} (\O_{X},\cE).
\end{align*}

\begin{Remark}\label{rem-1}
The classical derived category $\DC{X}$ is obtained from the classical
category $\Ch{X}$ of cochain complexes in 
$\Mod{X}$ by inverting the quasisomorphisms, and it is a triangulated
category.  The symbol $\heq$ therefore indicates an isomorphism
in the derived category.

These leads to some subtlely in viewing $\RHom_{X} (\cE,\cF)$ as the
morphisms of $\DC{X}$.  The problem is that, first of all, even to
build $\RHom_{X} (\cE,\cF)$ involves a choice of injective resolution
of $\cF$.  This choice interacts with the definition of the composition 
\begin{equation}\label{eq:1}
    \RHom_{X} (\cE,\cF) \times \RHom_{X} (\cF,\cG) \to 
    \RHom_{X} (\cE,\cG).
\end{equation}
Homological algebra implies that
$\RHom_{X} (\cE,\cF)$ is well-defined up to quasi-isomorphism, and so
the groups $\Ext_{X} (\cE,\cF)$ and the composition 
\[
\Ext_{X} (\cE,\cF) \times \Ext_{X} (\cF,\cG) \to 
    \Ext_{X} (\cE,\cG)
\]
are well-defined on the nose.  

For many purposes it is essential to deal with the category $\Ch{X}$ in
a way that recognizes the special importance of quasi-isomorphisms
without passing immediately to the derived category.  The theory of
$\i$-categories provides a convenient framework for doing so.  A key
feature of $\i$-categories is that they allow the cochain complex
$\Rhom_{X} (\cE,\cF)$ of morphisms to be defined only up to
quasi-isomorphism, and they recognize that the composition of
morphisms \eqref{eq:1} is necessarily an $A_{\infty}$ operation: the choices
involved in building an explicit model for the composition are
analogous to the choice of how to divide up the unit interval to
building a concatenation-of-paths operation 
\[
  \Map ([0,1],X) \times \Map ([0,1],X) \to   \Map ([0,1],X).
\]
The upshot is an $\i$-category $D^{b}_{\infty} (X)$ of chain complexes
of $\O_{X}$-modules, whose ``homotopy category'' is the
classical triangulated derived category $\DC{X}$ of Verdier.  
J. Lurie develops the theory of the derived category from this point
of view in \cite{lurie-dag-i}.  

In this paper, when we speak of the derived category we often
implicitly mean the $\i$-category $D^{b}_{\infty} (X)$ associated to
$\Mod{X}$:  For 
example, when we speak of $\RHom_{X} (\cE,\cF)$ as the morphisms in
the derived category, we really mean the $\i$-category.  When we use
the symbol $\heq$, we are indicating an equivalence in the
$\i$-category which becomes an isomorphism in $\DC{X}$. 

This $\i$-category plays an prominent role in the current mathematical
study of topological field theories.  For example, Costello 
\cite[Section 2.2]{costello} emphasizes that it is essential to use
$D^{b}_{\infty} (X)$ to obtain the $B$-model in his framework.
\end{Remark}

Associated to a map  $f: X \to Y$ of ringed spaces we 
have operations 
\[
\xymatrix{ 
           {\DC{Y}} \ar[r]|{{\bf L}f^{*}} &
           {\DC{X}} \ar@<1.5ex>[l]^{{\bf R}f_{*}}
                     \ar@<-1.5ex>[l]_{f_{!}}
}
\]
By construction, these operations satisfy properties analogous to
those for $f_{*}$ and $f^{*}$ above:
\begin{align*}
     \Arr f_{*}\cE \: & \heq \: \Arr\Gamma \cE \text{ if } Y \text{ is a one-point space}\\  
     \RHom_{X} ({\bf L}f^{*}\cE,\cF) \: & \heq \: \RHom_{Y} (\cE,{\bf R}f_{*}\cF)\\
     \Rhom_{X} (\cE,{\bf L}f^{*}\cG)  \: & \heq \: \Rhom_{Y} (f_{!}\cF,\cG)\\
\end{align*}
In particular, if $f: X \to \ptspace,$ then 
\[
    \Arr^{i}f_{*}\cE \: \iso \: H^{i} (X;\cE).
\]
We recall a number of additional properties enjoyed by these operations.

\subsubsection*{The projection formula}

Via their tensor products $\DC{X}$ and $\DC{Y}$ can be thought of
as categorical versions of commutative rings\footnote{
More generally, symmetric monoidal categories are categorical analogues of
commutative rings.  The category of ${\cal O}_X$ modules is a symmetric
monoidal category, and, because of the tensor products mentioned above,
the derived category $D^b(X)$ can also be given the structure of
a symmetric monoidal category.
}.
The pull-back $Lf^{*}$ preserves the tensor product, 
and so gives $\DC{X}$ in this sense the structure of a module over
$\DC{Y}.$ The \emph{projection formula} says that $\Arr f_{*}$  is a
homomorphism of modules over $\DC{Y}:$ for $\cE\in \DC{X}$ and $\cF\in
\DC{Y}$, we have 
\begin{equation}\label{eq:2}
       \Arr f_{*} (\cE \lotimes_{X}  \Ell f^{*}\cF) \: \heq \:  ( \Arr
       f_{*}\cE)\lotimes_{Y} \cF. 
\end{equation}

\subsubsection*{Flat base change}

Suppose that 
\[
\begin{CD}
X' @>g' >> X \\
@V f' VV @VV f V \\
Y' @> g >> Y.
\end{CD}   
\]
is a pull-back diagram: that is, $f':X' \to Y'$ is obtained from $f:
X\to Y$ by base change along $g: Y'\to Y$.

\begin{Proposition}[\cite{hart-ag}, prop. III.9.3; \cite{andrei-utah},
section 2.7]
 \label{flat-base-change}
If $g$ is flat, then so is $g'$, and so $\Ell g^{*} = g^{*}$ and $(\Ell
g')^{*} = (g')^{*}$.  Moreover
\[
     g^{*}\Arr f_{*} \heq ( \Arr f')_{*} (g')^{*}: \DC{X} \to \DC{Y}.
\] 
\end{Proposition}

\subsubsection*{Grothendieck-Serre duality}

The Grothendieck Duality theory describes the relationship between the
derived dual and the derived pushforward.  If $X$ is a smooth projective
variety of dimension $d$, let $S_{X} = \Omega^{d}_{X}[d] = K_{X}[d]$ be its
``dualizing complex.''   If $\cE$ is a complex of coherent (=finitely
generated quasi-coherent)
$\O_{X}$-modules, 
let 
\[
        \cE^{D} \: = \: \cE^{\vee}\lotimes_{X} S_{X} \: \heq \:
	\cE^{\vee}\otimes_{X}S_{X} \: \heq \: \shfRhom_{X} (\cE,S_{X}). 
\]
(if $X$ is smooth then $\Omega^{d}_{X}$ is projective, and so
$\lotimes$ and $\otimes$ coincide). 
If $f: X\to Y$ be a proper map of smooth projective varieties, then
the duality theory says that\footnote{We learned this formulation of
Grothendieck duality from A.~Neeman and A.~Caldararu.}
\begin{equation} \label{eq:5}
       {\bf R}f_{*} (\cE^{D}) \: \heq \: ({\bf R}f_{*}\cE)^{D}.
\end{equation}

\begin{Remark}
More generally, the Grothendieck Duality theory concerns the
existence and properties of a \emph{right} adjoint $f^{!}$ of the
functor $f_{*}$, so that, for $\cE\in \DC{X}$ and $\cF\in \DC{Y}$,
\begin{equation}\label{eq:3}
     {\bf R}f_{*}\shfRhom_{X} (\cE,f^{!}\cF)  \: \heq  \:
\shfRhom_{Y} ({\bf R}f_{*}\cE,\cF).
\end{equation}
For example, the theory asserts that if $f$ is proper and smooth of dimension
$d$, then  
\[
    f^{!}\cF \: \heq \: (f^{*}\cF)\lotimes \Omega^{d}_{X/Y}[d] \: \heq \:
    (f^{*}\cF)\otimes \Omega^{d}_{X/Y}[d].
\]
If $f:X\to Y$ is such that $f^{!}$ exists and is of the form 
\[
    f^{!}\cF \: \heq \: (f^{*}\cF)\lotimes S_{X/Y}
\]
for some element $S_{X/Y}$ of $\DC{X}$, then $S_{X/Y}$ is called a
\emph{dualizing complex} for the map $f.$  If $X$ and $Y$ are
themselves smooth, then 
\begin{equation}  \label{rel-dualizing}
   S_{X/Y} \: \heq \: S_{X}\otimes f^{*}S_{Y}^{-1}.
\end{equation}
Assuming that $\cE$ is finitely generated, and taking $\cF=\O_{Y}$,
\eqref{eq:3} becomes
\[
    {\bf R}f_{*} (\dual{\cE}\otimes S_{X}\otimes f^{*}S_{Y}^{-1}) \: \heq \: 
({\bf R}f_{*}\cE)^{\vee}.
\]
Using the projection formula \eqref{eq:2}, this becomes 
\[
    {\bf R}f_{*} (\dual{\cE}\otimes S_{X}) \otimes S_{Y}^{-1}) \: \heq \:
    ({\bf R}f_{*}\cE)^{\vee},
\]
so ${\bf R}f_{*} (\cE^{D}) \heq ({\bf R}f_{*}\cE)^{D}$, 
as asserted at \eqref{eq:5}.
\end{Remark}

\begin{Example}
Consider for example the case that $X$ is compact and smooth
and $Y=\ptspace$ is a the one-point space.  Then $S_{Y}=\C$,
considered as a sheaf over a point, and \eqref{eq:5} becomes
\[
   \Rhom_{X} (\cE,K_{X}[d]) \: \heq  \:
   \hom_{Y} ({\bf R}\Gamma \cE,\C).
\]
Taking cohomology, we find that 
\[
   \Ext^{d-i}_{X} (\cE,K_{X}) \: \iso \:
   \Hom (H^{i} (X;\cE),\C),
\]
which is Serre duality. 
\end{Example}

\subsection{Comparison of derived and underived operations}

\label{sec:prelim-c}

In this paper, our arguments for taffy identities in the B model
will technically revolve
around manipulations in derived categories, where it becomes
very simple to make strong statements.
However,
in many cases of physical interest, one really is using nonderived
operations on sheaves, not derived operations.
In this section, we will compare derived and underived operations,
to uncover the physics hidden in the technology of derived
categories. 

In particular in this section we study derived operations in the special case
that the derived category objects are pushforwards along embeddings of
vector bundles of finite rank, which is the simplest description of a
single set of D-branes. We shall discuss the relation between 
\begin{enumerate}
\item Derived pushforwards $\Arr p_{*}$ and ordinary pushforwards
$p_{*}$
\item Derived pullbacks ${\bf L} p^*$ and ordinary pullbacks
$p^*$
\item Derived duals $\vee$ and ordinary duals $*$
\item Derived tensor products $\otimes^{\bf L}$
and ordinary tensor products $\otimes$
\end{enumerate}
for such special objects in derived categories.
We shall see, for example, that the Freed-Witten anomaly \cite{freed-ed} is
implicit in derived duals.

Before specializing to pushforwards of vector bundles,
we mention in passing
the simplest relationships of the derived operations to their underived
analogues: 
\begin{enumerate}
\item $\Arr f_{*}E\heq f_{*}E$ if $E$ is injective or if $f$ is affine
({\it e.g.} a
closed embedding); 
\item $\Ell f^{*}E\heq f^{*}E$ if $E$ is projective or if $f$ is flat;
\item $\Rhom_{X} (E,F) \heq \hom_{X} (E,F)$ if either $E$ is projective or $F$
is injective;
\item $E\otimes_{X} F \heq E \lotimes_{X} F$ if either $E$ or $F$ is
projective ({\it e.g.} free).
\end{enumerate}

\subsubsection*{Derived pushforwards}

In the B model we can consider a $D$-brane on a closed submanifold
$i: S \to X$ described
by the sheaf $i_* E$, with $E$ a finite rank vector bundle over $S$.
Since $i$ is a
closed embedding, it follows that 
\[
   \Arr i_{*}E \heq i_{*}E
\]
(and indeed for any sheaf on $S$).  We shall make considerable
use of the particular case 
\[
  \Delta: X \to X\times X
\]
of the diagonal embedding: $\Arr\Delta_{*} \heq \Delta_{*}$.

\subsubsection*{Derived pullbacks}

In this paper we shall be considering primarily pullbacks along
projections,  {\it e.g.} $p: X \times Y \rightarrow X$.  Projections
are flat, and so derived pullbacks match ordinary pullbacks.

In addition, we sometimes use pullbacks along other maps in
derivations.
However, in other cases, derived and ordinary pullbacks typically
do not match.
For example, we occasionally discuss pullbacks along closed embeddings,
such as the diagonal map $\Delta: X \rightarrow X \times X$.  
In the special case of pullbacks of vector bundles,
the derived and ordinary pullbacks (along closed embeddings) do match.
On the other hand, for more general sheaves, the derived and ordinary
pullbacks do not match, essentially because closed embeddings are
not flat.  For example, $\Delta^* {\cal O}_{\Delta} = 
{\cal O}_X$; however, ${\bf L} \Delta^* {\cal O}_{\Delta}
\neq {\cal O}_X$, as there are nonzero cohomology sheaves
(roughly $\Omega^i$ in position $-i$).

\subsubsection*{Derived duals}

Given a vector bundle of finite rank $E$ over a submanifold 
$i: S \hookrightarrow
X$ of codimension $d$, consider the
derived dual 
\begin{displaymath}
({\bf R}i_* E)^{\vee} \: \equiv \:
{\bf R}{\it Hom}( \Arr i_* E, {\cal O}_X ) \: \heq \: 
{\bf R}{\it Hom}(i_* E, {\cal O}_X ).
\end{displaymath}
(As we noted above, $\Arr i_{*} \heq i_{*}$, as $i$ the inclusion of a
submanifold.)  We shall see that
it is related to $i_* (E^{*})$ by a combination of
grading shifts and tensoring with canonical bundles.  
Specifically, 
\begin{equation}\label{eq:4}
    (i_{*}E)^{\vee} \: \heq \: {\bf R} i_{*} (E^{*}\otimes S_{S/X})
\end{equation}
where $S_{S/X}$ is the dualizing sheaf of the embedding, and $*$ denotes
the ordinary vector bundle dual.
To see this, note that, using \eqref{eq:5}, we have
\[
   (i_{*}E)^{D} \: \heq  \: i_{*} (E^{D}).
\]
That is, 
\begin{align*}
     (i_{*}E)^{\vee} \: & \heq \: (i_{*}E)^{D}\otimes S_{X}^{-1} \\
\: & \heq \: i_{*} (E^{D})\otimes S_{X}^{-1} \\
\: & \heq \: i_{*} (E^{\vee} \otimes S_{S} \otimes i^{*}S_{X}^{-1}) \\
\: & \heq \: {\bf R} i_{*} (E^{*}\otimes S_{S/X}).
\end{align*}
In the preceding, we used the fact that $X$ and $S$ are smooth, so
$S_{S}$ and $S_{X}$ are projective: thus derived tensor products and
pull-backs reduce to ordinary ones.  We used the fact that $E$
is a vector bundle, and so locally free, to conclude that 
\[
    \dual{E} = \shfRhom_{X} (E,\O_{X}) \: \heq \: \shfhom_{X} (E,\O_{X}) 
\: = \: E^{*}.
\]
We used the projection formula to conclude that 
\[
    i_{*} (\cF) \lotimes S_{X}^{-1}
\: \heq \: i_{*} (\cF\lotimes {\bf L}i^{*} S_{X}) \: \heq \:
 i_{*} (\cF\otimes i^{*}S_{X}).
\]

It remains to compute $S_{S/X}$.
From equation~(\ref{rel-dualizing}), we know
that 
\begin{displaymath}
S_{S/X}\: = \: S_{S}\otimes i^{*}S_{X}^{-1} \: = \:
K_S[{\rm dim}\: S] \otimes i^* \left( K_X[{\rm dim}\: X] \right)^{-1}
\: = \: K_S \otimes \left( K_X|_S \right) [ - d ]
\end{displaymath}
where recall $d$ is the codimension of $S$ in $X$. 
For completeness, note from the
short exact sequence  
\[
   0 \: \longrightarrow \: N_{S/X}^{*} \: \longrightarrow \:
   \Omega^{1}_{X}|_{S} \: \longrightarrow \: \Omega^{1}_{S} 
\: \longrightarrow \:  0
\]
it is straightforward to show that
\begin{displaymath}
\Lambda^{\rm top} N_{S/X} \: \cong \:
K_S \otimes \left( K_X|_S \right)^{-1}
\end{displaymath}
hence
\[
     S_{S/X} \: = \: \Lambda^{\rm top} N_{S/X} [ - d].
\]

Putting this together, we have that
\begin{equation}   \label{der-dual-final}
\left( i_* E \right)^{\vee} \: \cong \:
i_*\left( E^* \otimes K_S \otimes \left( K_X|_S \right)^{-1}
\right)
[ - d ]
\: \cong \:
i_*\left( E^* \otimes \Lambda^{\rm top} N_{S/X}
\right)
[ - d ].
\end{equation}

\begin{Example}
Consider the diagonal embedding $\Delta: X \to X\times
X$.  Then $N_{X/X\times X} \iso T_{X}$,  
$K_{X \times X} = \pi_1^* K_X \otimes \pi_2^* K_X$,
$K_{X \times X}|_X =  K_X^{\otimes 2}$,
and \eqref{der-dual-final} becomes
\begin{equation}\label{eq:6}
    (\Delta_{*} {\cal O}_{X})^{\vee} \: \heq \: \Delta_{*}  \left(
{\cal O}_X \otimes K_X \otimes \left( K_X^{\otimes 2} \right)^{-1} \right) 
[-{\rm dim} \: X] 
\: = \: \Delta_{*}K_{X}^{-1}[-{\rm dim}\: X].
\end{equation}
(See also \cite{aw} for a discussion of the dual of the diagonal.)
\end{Example}

Now, let us return to our discussion of the derived dual
of $i_* E$.  In this case, for $X$ a Calabi-Yau,
equation~(\ref{der-dual-final}) becomes
\begin{displaymath}
\left( i_* E \right)^{\vee} \: = \:
i_*\left( E^* \otimes K_S \right)[ -d ].
\end{displaymath}

Note that the factor of $K_S$ is ultimately due to the
Freed-Witten anomaly \cite{freed-ed}.
Recall from \cite{ks} that because of the Freed-Witten
anomaly, the sheaf $i_* E$ is associated with D-brane
Chan-Paton 
factors\footnote{
It has been suggested by A.~Caldararu that an alternative approach to the
Freed-Witten anomaly would be, in part, to define a ``good'' dual
to a sheaf ${\cal E}$ by,
${\cal E}^* \equiv {\cal E}^{\vee} \otimes \sqrt{S_X}$, when this makes
sense.  We shall not follow that path here, but thought it warranted
mentioning.
} $E \otimes K_S^{-1/2}$.  If we dualize the Chan-Paton
factors, they become 
\[
\left( E \otimes K_S^{-1/2} \right)^* \: = \:
E^* \otimes K_S^{+1/2} = E^* \otimes
K_{S}^{-1/2} \otimes K_S;
\]
which means that the appropriate dual of the sheaf
$i_* E$ (modulo grading shifts) should be
$i_{*} (E^{*}\otimes K_{S}) \heq
i_{*} (E^{D}) \heq (i_{*} E)^{D}.$ 
In effect, this means that the Freed-Witten anomaly is baked into the
formalism of derived categories, as it is automatically encoded
in the natural dual in the sense of
derived categories (the derived dual).

\subsubsection*{Derived tensor products}

Next, let us compare the derived tensor product
$\otimes^{\bf L}$ to the ordinary tensor product
$\otimes$.   Consider two D-branes wrapped on $i: S \hookrightarrow X$ and
$j: T \hookrightarrow X$, $S$, $T$, and $X$ all assumed smooth.
If the intersection of $S$ and $T$ is transversal
({\it e.g.} 
${\rm codim}\, S/X + {\rm codim}\, T/X = {\rm codim}\, (S \cap T)/X$),
then it is a result of Serre that
\begin{displaymath}
\left( i_* E \right) \otimes^{\bf L} \left( j_* F \right)
\: \heq \:
\left( i_* E \right) \otimes \left( j_* F \right).
\end{displaymath}

(The point is to show that the sheaf $\uln{\Tor}^{p} (i_{*}E,j_{*}F)$
is zero for $p>0$, which is precisely when the derived and underived
tensor products will match.  One reduces to the case that the ring of functions
on $X$ is $A$, on $S$ is $A/I$, and $T$ is $A/J$ for prime ideals $I$
and $J.$  With our hypotheses $\Tor^{A} (A/I,A/J)$ can be computed by
a Koszul complex, and the transverse intersection assumption implies
that this complex is acylic.  The relevant results are p. 54 Prop 2,
and p. 55 Cor 2 of \cite{serre}.)

More generally, however, for non-transversal intersections, the derived
tensor product will differ from the ordinary tensor product.
Let $k: S \cap T \hookrightarrow X$ denote the natural embedding of the
intersection, then the extra contributions arise from
\begin{displaymath}
k_* \left( E|_{S \cap T} \otimes F|_{S \cap T} \otimes
\Lambda^* B \right)
\end{displaymath}
where
\begin{displaymath}
B^* \: = \: TX|_{S \cap T} / \left( TS|_{S \cap T} + TT|_{S \cap T} 
\right)
\end{displaymath}
(a bundle which also appeared in \cite{ks}).
The bundle $B$ expresses the amount by which the intersection fails
to be transversal.

In this paper, we shall only need the simple transverse case.  
In every case in which we are initially reading off string states
from a diagram, the sheaves will be supported on different factors in a 
product.
Let $E$ be a vector bundle on $i: S \hookrightarrow X$,
and $F$ be a vector bundle on $j: T \hookrightarrow Y$.
Then the ordinary and derived tensor products match \cite{andreipriv0}:
\begin{displaymath}
p_1^* i_* E \otimes p_2^* j_* F \: = \:
p_1^* i_* E \otimes^{\bf L} p_2^* j_* F
\end{displaymath}
and if we let $k$ denote the inclusion $S \times T \hookrightarrow X
\times Y$, then they both match
\begin{displaymath}
k_* \left( p_1^* E \otimes p_2^* F \right)
\end{displaymath}

In the bulk of the rest of this paper, we will study examples of
taffy-like foldings of string worldsheets, and derive mathematical
identities to check that such taffy operations give alternative descriptions
of physically-equivalent string worldsheets.

Our constructions in the B model necessarily take place in the 
derived category, and as
such, all operations will necessarily be derived.
For notational simplicity, in the rest of this paper after this subsection, 
we will use underived notation to implicitly mean derived constructions.
In other words,
we will omit the $\Arr$ and
the $\Ell$ from the notation: thus $\Hom$ means
$\Rhom$, $\otimes$ means $\lotimes$, $i_{*}$ means $\Arr i_{*}$, 
and so forth.

\subsection{Open string states}

Let ${\cal E}$, ${\cal F}$ be objects in $D^b(X)$, defining boundaries
in the open string B model on $X$.

Here is our first example of a taffy operation.
Consider folding an oriented open string 
\begin{center}
\begin{picture}(100,20)
\ArrowLine(5,10)(95,10)
\Vertex(5,10){2}  \Vertex(95,10){2}
\Text(0,10)[r]{${\cal E}$}
\Text(100,10)[l]{${\cal F}$}
\end{picture}
\end{center}
along a trivial defect inserted at the center
into a U-shape:
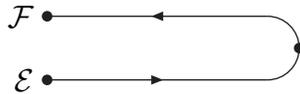
\begin{center}
\begin{picture}(100,50)
\ArrowLine(5,12)(88,12)    \ArrowLine(88,36)(5,36)
\CArc(88,24)(12,-90,90)  \Vertex(100,24){2}
\Vertex(5,12){2}   \Vertex(5,36){2}
\Text(0,12)[r]{${\cal E}$}
\Text(0,36)[r]{${\cal F}$}
\end{picture}
\end{center}
We mean by the diagram above (and other such in this paper)
to indicate that the original open string
has been folded over and then collapsed onto a single new open string on
$X \times X$, with one boundary determined by ${\cal E}$ and ${\cal F}$,
and the other boundary determined by the identity defect,
as represented by a diagonal embedding
$\Delta: X \rightarrow X \times X$:
\begin{center}
\begin{picture}(100,20)
\Line(5,10)(95,10)
\Vertex(5,10){2}  \Vertex(95,10){2}
\Text(0,10)[r]{${\cal E}, {\cal F}$}
\Text(100,10)[l]{$\Delta$}
\end{picture}
\end{center}
In order to convey more information, the previous diagram was
`expanded' vertically, to show the different layers of the
original open string, though the reader should always interpret
such diagrams to mean that a vertical contraction onto a single-layer
open string on multiple copies of $X$ has taken place.

In effect, we are encoding worldsheet geometry in D-branes.
See also \cite{ed-str} where something analogous was done in
a different theory.

We are not quite done.  We need to uniquely specify Chan-Paton factors
on the folded string ({\it e.g.} ${\cal F}$ versus ${\cal F}^{\vee}$),
and we also need to specify an orientation on the folded open string.
It is straightforward to see that
different reasonable choices differ only by grading shifts, so this
is a matter of picking a convention, not something essential to the physics.
We shall follow the following convention (which was chosen specifically
to preserve gradings):
\begin{enumerate}
\item For diagonals, if the orientation on the folded string points
{\it towards} the diagonal, we describe the Chan-Paton factors by
the object $\Delta_* {\cal O}_X$ in the derived category,
which (because it will appear commonly) we shall abbreviate
$\Delta$.
If the orientation on the folded string points {\it away} from the
diagonal, we use instead the object $\Delta^{\vee}$.
(Recall from equation~(\ref{eq:6}) that on a Calabi-Yau, 
$\Delta$ differs from $\Delta^{\vee}$ merely by a grading shift, 
$\Delta^{\vee} = \Delta[- {\rm dim}\, X]$, so we are merely choosing
conventions so as to preserve gradings.)
\item For other objects present before the fold (in the example above,
${\cal E}$, ${\cal F}$), if the orientation on the folded string is
parallel to the orientation on the part of the original string going
into that boundary, we use the original object; if antiparallel,
we use its dual.
\end{enumerate}

For example, if we pick the orientation on the folded string to be
\begin{center}
\begin{picture}(100,20)
\ArrowLine(5,10)(95,10)
\Vertex(5,10){2}  \Vertex(95,10){2}
\Text(0,10)[r]{${\cal E}, {\cal F}$}
\Text(100,10)[l]{$\Delta$}
\end{picture}
\end{center}
then the left Chan-Paton factors are described by the derived category object 
$\pi_1^* {\cal E} \otimes \pi_2^* {\cal F}^{\vee}$,
and the right Chan-Paton factors are described by the derived category object
$\Delta$.
In this case, the string states in the new (folded) open string
are counted by
\begin{displaymath}
{\rm Ext}^*_{X \times X} \left( \pi_1^* {\cal E} \otimes
\pi_2^* {\cal F}^{\vee}, \Delta \right)
\end{displaymath}
Applying the identities from section~\ref{prelims-r}, we have
\begin{eqnarray*}
\lefteqn{
{\rm {\bf R}Hom}_{X \times X}( \pi_1^* {\cal E} \otimes_{X\times X}   \pi_2^* {\cal F}^{\vee}, 
\Delta_* {\cal O}_X)
} \\
& \heq &
{\rm {\bf R}Hom}_X(\Delta^* (\pi_1^* {\cal E} \otimes_{X\times X}  \pi_2^* {\cal F}^{\vee}), 
{\cal O}_X) \\
& \heq &
{\rm {\bf R}Hom}_X({\cal E}
\otimes {\cal F}^\vee, {\cal O}_X) \: \heq\: {\rm {\bf R}Hom}_X({\cal E}, {\cal F}).
\end{eqnarray*}
and taking cohomology yields the equality
\begin{equation}   \label{fold-strip-U}
{\rm Ext}^*_{X \times X} \left( \pi_1^* {\cal E} \otimes 
\pi_2^* {\cal F}^{\vee},
\Delta \right)
\: = \:
{\rm Ext}^*_X\left( {\cal E}, {\cal F} \right) 
\end{equation}
so we see that in this convention, the open string states on the
folded string precisely match the original open string states, 
without even a grading shift.

If we had picked the opposite orientation on the folded string:
\begin{center}
\begin{picture}(100,20)
\ArrowLine(95,10)(5,10)
\Vertex(5,10){2}  \Vertex(95,10){2}
\Text(0,10)[r]{${\cal E}, {\cal F}$}
\Text(100,10)[l]{$\Delta$}
\end{picture}
\end{center}
then the left Chan-Paton factors would be described by the
derived category object $\pi_1^* {\cal E}^{\vee} \otimes
\pi_2^* {\cal F}$, and the right Chan-Paton factors would be
described by the derived category object $\Delta^{\vee}$.
In this case, the open string states in the folded open string
would be given by
\begin{displaymath}
{\rm Ext}^*_{X \times X}\left( \Delta^{\vee}, 
\pi_1^* {\cal E}^{\vee} \otimes \pi_2^* {\cal F} \right)
\end{displaymath}
but a trivial application of section~\ref{prelims-r} and 
identity~(\ref{fold-strip-U}) implies
\begin{displaymath}
{\rm Ext}^*_{X \times X}\left( \Delta^{\vee}, 
\pi_1^* {\cal E}^{\vee} \otimes \pi_2^* {\cal F} \right)
\: = \:
{\rm Ext}^*_{X \times X}\left(
\pi_1^* {\cal E} \otimes \pi_2^* {\cal F}^{\vee}, \Delta \right)
\: = \:
{\rm Ext}^*_X\left( {\cal E}, {\cal F} \right)
\end{displaymath}
so again we recover the states of the original open string,
without even a grading shift.

One technical point is that in the description above, the diagonal defect
$\Delta$ is implicitly assumed to have (trivial, rank 1) Chan-Paton factors.
The fact that the boundaries have support along the diagonal
corresponds to the fact that they correspond to folds, but, 
folds do not have Chan-Paton
factors.  However, the
open string spectrum with trivial line bundles along the defect
$\Delta: X \rightarrow X \times X$ is the same as the open string
spectrum with no Chan-Paton factors added at all (a trivial consequence
of {\it e.g.} \cite{ks}).  Thus, we can equivalently interpret the Ext
groups above as computing open string spectra between boundaries with
no added Chan-Paton factors.

Now, let us perform a consistency check of our description of folded
open strings.  Consider the folded open string with orientation
\begin{center}
\begin{picture}(100,20)
\ArrowLine(5,10)(95,10)
\Vertex(5,10){2}  \Vertex(95,10){2}
\Text(0,10)[r]{${\cal E}, {\cal F}$}
\Text(100,10)[l]{$\Delta$}
\end{picture}
\end{center}
so that the string states are
\begin{displaymath}
{\rm Ext}^*_{X \times X} \left( \pi_1^* {\cal E} \otimes
\pi_2^* {\cal F}^{\vee}, \Delta \right)
\: = \: 
{\rm Ext}^*_X\left( {\cal E}, {\cal F} \right)
\end{displaymath}
Now that we have fixed an orientation on the folded open string corresponding
to the orientation on the original string, we can ask about the effect
of flipping the orientation.  If we were to traverse the folded string
in the opposite direction, the string states would be
\begin{eqnarray*}
{\rm Ext}^*_{X \times X}\left( \Delta,
\pi_1^* {\cal E} \otimes \pi_2^* {\cal F}^{\vee} \right)
& = &
{\rm Ext}^*_{X \times X}\left( 
\pi_1^* {\cal E}^{\vee} \otimes \pi_2^* {\cal F}, \Delta^{\vee} \right) \\
& = & {\rm Ext}^*_{X \times X}\left(
\pi_1^* {\cal E}^{\vee} \otimes \pi_2^* {\cal F}, \Delta [ - {\rm dim}\, X]
\right) \: \mbox{ (for $X$ Calabi-Yau)} \\
& = & {\rm Ext}^*_X\left( {\cal E}^{\vee}, {\cal F}^{\vee}[ - {\rm dim}\, X]
\right) \\
& = & {\rm Ext}^*_X\left( {\cal F}, {\cal E} [ - {\rm dim}\, X] \right)
\end{eqnarray*}
which, up to an irrelevant grading shift, are precisely the states
one would get from traversing the original unfolded string in the opposite
direction, as expected.

For completeness, let us also consider folding in the opposite direction:
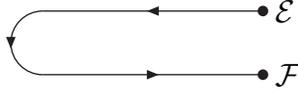
\begin{center}
\begin{picture}(100,50)
\ArrowLine(12,12)(95,12)
\ArrowLine(95,36)(12,36)
\ArrowArc(12,24)(12,90,270)
\Vertex(95,12){2}  \Vertex(95,36){2}
\Text(100,12)[l]{${\cal F}$}
\Text(100,36)[l]{${\cal E}$}
\end{picture}
\end{center}
If we collapse this to an open string on $X \times X$ with the following
orientation:
\begin{center}  
\begin{picture}(100,20)\ArrowLine(95,10)(5,10)\Vertex(5,10){2}  \Vertex(95,10){2}
\Text(100,10)[l]{${\cal E}, {\cal F}$}
\Text(0,10)[r]{$\Delta$}
\end{picture}
\end{center}
then following our usual convention, string states are given by
\begin{displaymath}
{\rm Ext}^*_{X \times X}\left( \pi_1^* {\cal E} \otimes 
\pi_2^* {\cal F}^{\vee}, \Delta \right)
\end{displaymath}
and we have already seen that this matches
${\rm Ext}^*_X\left( {\cal E}, {\cal F} \right)$.
Similarly for the opposite orientation convention on the folded string.

In order to show that similar results apply for more complicated
folds,
we will need a slight generalization of the
identity~(\ref{fold-strip-U}), whose proof we learned from
\cite{andreipriv0}.  For  any $\cE\in D^{b} (X)$ and any finitely
generated ${\cal S} \in
D^{b} (X \times Y)$, we have 
\begin{equation}    \label{U-shape-main}
{\rm Ext}^*_{X \times X \times Y} \left(
\pi_1^* {\cal E} \otimes \pi_{23}^* {\cal S}^{\vee},
\pi_{12}^* \Delta \right)
\: = \:
{\rm Ext}^*_{X \times Y}\left( \pi_1^* {\cal E},
{\cal S} \right)
\end{equation}
This is our first ``taffy identity.''
Intuitively, it allows us to equate the two diagrams
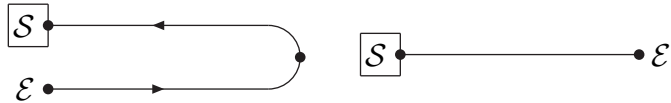
\begin{center}
\begin{picture}(120,50)
\ArrowLine(15,12)(98,12)    \ArrowLine(98,36)(15,36)
\CArc(98,24)(12,-90,90)  \Vertex(110,24){2}
\Vertex(15,12){2}   \Vertex(15,36){2}
\Text(10,12)[r]{${\cal E}$}
\Text(10,36)[r]{${\cal S}$}
\Line(0,44)(15,44)  \Line(15,44)(15,28)  \Line(15,28)(0,28)  \Line(0,28)(0,44)
\end{picture}
$\:  \:$
\begin{picture}(120,50)
\Line(15,25)(105,25)
\Vertex(15,25){2}   \Vertex(105,25){2}
\Line(0,33)(15,33) \Line(15,33)(15,17)  \Line(15,17)(0,17)  \Line(0,17)(0,33)
\Text(10,25)[r]{${\cal S}$}
\Text(110,25)[l]{${\cal E}$}
\end{picture}
\end{center}
(where ${\cal S}$ is being used to encode any number of additional
foldings)
and hence can be used to unfold foldings.

To prove~(\ref{U-shape-main}), note first of all that 
\[
{\rm Ext}^*_{X \times X \times Y}  \left(
\pi_1^* {\cal E} \otimes \pi_{23}^* {\cal S}^{\vee},
\pi_{12}^* \Delta \right)
\: = \: 
\Ext^{*}_{X\times X \times Y}  \left(
\pi_1^* {\cal E},\pi_{23}^* {\cal S} \otimes 
\pi_{12}^* \Delta \right)
\]
Let 
\[
\Delta_{12}: \: (x,y) \: \mapsto \: (x,x,y), \: \: \:
\Delta: \: x \:  \mapsto \: (x,x).
\]
We have the pull-back diagram 
\begin{displaymath}
\xymatrix{
X \times Y \ar[r]^-{ \pi_1 } \ar[d]^{ \Delta_{12} } & X \ar[d]^{\Delta} \\
X \times X \times Y \ar[r]^-{\pi_{12} } & X \times X
}
\end{displaymath}
with $\pi_{12}$ flat, and so by flat base change (prop.~\ref{flat-base-change})
we have 
\[
          \pi_{12}^{*}\Delta_{*}\O_{X} \heq
	  \Delta_{12*}\pi_{1}^{*}\O_{X} = \Delta_{12*}O_{X\times Y}.
\]
Now using the projection formula (equation~(\ref{eq:2})) and the fact that
$\pi_{23}\circ \Delta_{12} (x,y) = (x,y),$ we have 
\begin{align*}
      \pi_{23}^{*}\mathcal{S}\otimes \pi_{12}^* \Delta & \heq 
      \pi_{23}^{*}\mathcal{S}\otimes \Delta_{12*}O_{X\times Y} \\
&    \heq       \Delta_{12*} (\Delta_{12}^{*}\pi_{23}^{*}\mathcal{S})  \\
& \heq \Delta_{12*}\mathcal{S}.
\end{align*}
Thus (noting that $\pi_{1}\circ \Delta_{12} = \pi_{1}$)
\begin{align*}
\Ext^*_{X \times X \times Y}  \left(
\pi_1^* {\cal E} \otimes \pi_{23}^* {\cal S}^{\vee},
\pi_{12}^* \Delta \right)
\: & = \: 
\Ext^{*}_{X\times X \times Y}  \left(
\pi_1^* {\cal E},\Delta_{12*}\mathcal{S} \right) \\
& = \: 
\Ext^{*}_{X}  \left(\cE, \pi_{1*}\Delta_{12*}\mathcal{S}\right)  \\
& = \: 
\Ext^{*}_{X}  \left(\cE, \pi_{1*}\mathcal{S}\right)  \\
& = \: 
\Ext^{*}_{X\times Y}  \left(\pi_{1}^{*}\cE, \mathcal{S}\right),
\end{align*}
which is \eqref{U-shape-main}.

One trivial application of the result above is to 
recheck the U-shaped-strings discussed
previously.
There, we argued that
\begin{displaymath}
{\rm Ext}^*_{X \times X}\left( \pi_1^* {\cal E} \otimes
\pi_2^* {\cal F}^{\vee}, \Delta \right) \: = \:
{\rm Ext}^*_X\left( {\cal E}, {\cal F} \right)
\end{displaymath}
directly from homological algebra.
If instead we apply~(\ref{U-shape-main}), taking $Y$ to be a point and
${\cal S} = {\cal F}$, then we immediately recover the same result.

A more interesting example is to fold an ordinary open string
on $X$ with boundaries ${\cal E}$, ${\cal F}$ into an S-shape,
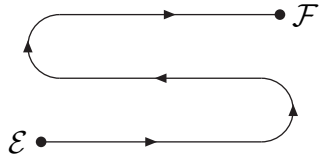
\begin{center}
\begin{picture}(100,80)
\ArrowLine(5,26)(88,26)
\ArrowLine(88,50)(12,50)
\ArrowLine(12,74)(95,74)
\ArrowArc(88,38)(12,-90,90)
\ArrowArcn(12,62)(12,-90,90)
\Vertex(5,26){2}  \Vertex(95,74){2}
\Text(0,26)[r]{${\cal E}$}
\Text(100,74)[l]{${\cal F}$}
\end{picture}
\end{center}
Pressing this down into a single open string on $X^3$, and picking
an orientation on the folded string, we recover
\begin{center}
\begin{picture}(100,20)
\ArrowLine(5,10)(95,10)
\Vertex(5,10){2}  \Vertex(95,10){2}
\Text(0,10)[r]{${\cal E}, \Delta_{23}$}
\Text(100,10)[l]{$\Delta_{12}, {\cal F}$}
\end{picture}
\end{center}
with states
\begin{displaymath}
{\rm Ext}^*_{X \times X \times X} \left(
\pi_1^* {\cal E} \otimes \pi_{23}^* \Delta^{\vee},
\pi_{12}^* \Delta \otimes \pi_3^* {\cal F} \right).
\end{displaymath}
In order for the folded string to be equivalent to the original,
we conjecture
\begin{equation}   \label{fold-strip-S}
{\rm Ext}^*_X\left( {\cal E}, {\cal F} \right) \: = \:
{\rm Ext}^*_{X \times X \times X} \left(
\pi_1^* {\cal E} \otimes \pi_{23}^* \Delta^{\vee},
\pi_{12}^* \Delta \otimes \pi_3^* {\cal F} \right).
\end{equation}
(See also {\it e.g.}
\cite{bachasetal}[fig. 6a] for a different discussion of this
same diagram.)

The identity above can be derived by repeatedly applying
identity~(\ref{U-shape-main}).  For notational convenience, define
\begin{displaymath}
\Delta_{ij} \: \equiv \: \pi_{ij}^* \Delta_* {\cal O}_X
\end{displaymath}
where $\pi_{ij}$ is the projection onto the $i$th, $j$th factors of
$X$ in a product of several copies.
Then the massless states associated to the S-shape above
are given by
\begin{align*}
{\rm Ext}^*_{X^3}\left( \pi_1^* {\cal E} \otimes \Delta_{23}^{\vee},
\Delta_{12} \otimes \pi_3^* {\cal F} \right) 
\:& = \:
{\rm Ext}^*_{X^3}\left( \pi_1^* {\cal E} \otimes \Delta_{23}^{\vee} \otimes
\pi_3^* {\cal F}^{\vee}, \Delta_{12} \right) \\
\:& = \: 
{\rm Ext}^*_{X^3}\left( \pi_1^* {\cal E} \otimes \pi_{23}^{*} (\Delta \otimes
\pi_2^* {\cal F})^{\vee}, \Delta_{12} \right) 
\end{align*}
Applying the identity~(\ref{U-shape-main}), 
which means, unrolling the bottommost 
U-shape, this becomes
\begin{displaymath}
{\rm Ext}^*_{X^2} \left( \pi_1^* {\cal E},
\Delta \otimes \pi_2^* {\cal F} \right)
\: = \:
{\rm Ext}^*_{X^2} \left(\pi_{1}^{*}\cE \otimes \dual{\cF}, \Delta\right), 
\end{displaymath}
or graphically
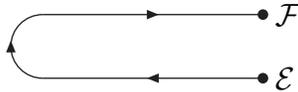
\begin{center}
\begin{picture}(100,50)
\ArrowLine(95,12)(12,12)
\ArrowLine(12,36)(95,36)
\ArrowArcn(12,24)(12,-90,90)
\Vertex(95,12){2}  \Vertex(95,36){2}
\Text(100,12)[l]{${\cal E}$}
\Text(100,36)[l]{${\cal F}$}
\end{picture}
\end{center}
We have already seen that the massless states in this diagram are just 
$\Ext_{X}^{*}(\cE,\cF)$, hence we recover~(\ref{fold-strip-S}).

The pattern can now be repeated {\it ad infinitum}, repeatedly
folding open strings over themselves to produce open strings on 
higher-dimensional spaces.  
To drive home that point,
let us work through one last example of these manipulations.
Consider refolding an open string over itself into a five-layer pattern,
as shown:
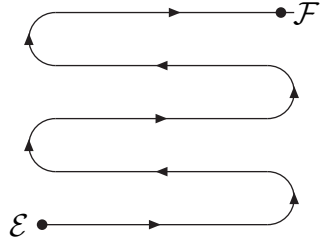
\begin{center}
\begin{picture}(100,100)
\ArrowLine(5,10)(90,10)
\ArrowLine(90,30)(10,30)
\ArrowLine(10,50)(90,50)
\ArrowLine(90,70)(10,70)
\ArrowLine(10,90)(100,90)
\Vertex(5,10){2}  \Vertex(95,90){2}
\ArrowArc(90,20)(10,-90,90)
\ArrowArcn(10,40)(10,-90,90)
\ArrowArc(90,60)(10,-90,90)
\ArrowArcn(10,80)(10,-90,90)
\Text(0,10)[r]{${\cal E}$}
\Text(100,90)[l]{${\cal F}$}
\end{picture}
\end{center}
The result of this folding is an open string on the product of
five copies of $X$, with states counted by
\begin{displaymath}
{\rm Ext}^*_{X^5}\left(
\pi_1^* {\cal E} \otimes \Delta_{23}^{\vee} \otimes \Delta_{45}^{\vee},
\Delta_{12} \otimes \Delta_{34} \otimes \pi_5^* {\cal F} \right)
\end{displaymath}
Since this is just a folded version of a string between ${\cal E}$ and
${\cal F}$, the states should be counted by ${\rm Ext}_X^*\left( {\cal E},
{\cal F} \right)$, which we shall now check by successive unfolding
operations.
We can begin by unfolding the bottommost U-shape,
using identity~(\ref{U-shape-main}),
revealing that the states above are the same as
\begin{displaymath}
{\rm Ext}^*_{X^4}\left(
\pi_1^*{\cal E}, \Delta_{12} \otimes \Delta_{34}
\otimes \Delta_{23} \otimes \pi_4^* {\cal F} \right)
\: = \: {\rm Ext}^*_{X^4}\left(
\pi_1^* {\cal E} \otimes \Delta_{12}^{\vee} \otimes \Delta_{34}^{\vee},
\Delta_{23} \otimes \pi_4^* {\cal F} \right)
\end{displaymath}
or graphically
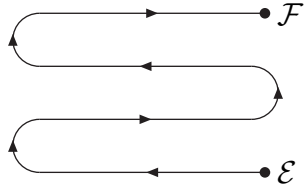
\begin{center}
\begin{picture}(100,90)
\ArrowLine(95,20)(10,20)
\ArrowLine(10,40)(90,40)
\ArrowLine(90,60)(10,60)
\ArrowLine(10,80)(95,80)
\Vertex(95,20){2}  \Vertex(95,80){2}
\ArrowArcn(10,30)(10,-90,90)
\ArrowArc(90,50)(10,-90,90)
\ArrowArcn(10,70)(10,-90,90)
\Text(100,20)[l]{${\cal E}$}
\Text(100,80)[l]{${\cal F}$}
\end{picture}
\end{center}
Unrolling the next bottom U-shape, applying identity~(\ref{U-shape-main}),
we see that the states above are the same as
\begin{displaymath}
{\rm Ext}^*_{X^3}\left(
\pi_1^* {\cal E} \otimes \Delta_{23}^{\vee}, \Delta_{12} \otimes
\pi_3^* {\cal F} \right)
\end{displaymath}
or graphically
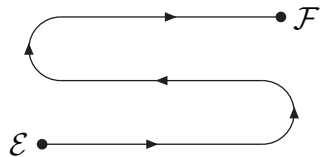
\begin{center}
\begin{picture}(100,100)
\ArrowLine(5,26)(88,26)
\ArrowLine(88,50)(12,50)
\ArrowLine(12,74)(95,74)
\ArrowArc(88,38)(12,-90,90)
\ArrowArcn(12,62)(12,-90,90)
\Vertex(5,26){2}  \Vertex(95,74){2}
\Text(0,26)[r]{${\cal E}$}
\Text(100,74)[l]{${\cal F}$}
\end{picture}
\end{center}
This is identical to the S-shape discussed earlier, so we can now
conclude that the states in this folded open string are indeed
counted by ${\rm Ext}^*_X({\cal E}, {\cal F})$, as expected.

It should be clear that this process can be continued for
arbitrarily many folds; regardless of the number of foldings
introduced, if we start with an open string from ${\cal E}$ to
${\cal F}$, then after foldings the open string states will still
be given by ${\rm Ext}^*_X({\cal E}, {\cal F})$.

\subsection{Closed string states}  \label{bmodel-closed}

Now, let us apply the same ideas to closed strings.
Begin with a closed string on $X$,
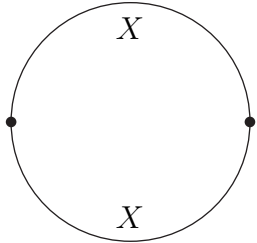
\begin{center}
\begin{picture}(100,100)(0,0)
\CArc(50,50)(45,0,180)
\CArc(50,50)(45,180,360)
\Vertex(95,50){2}
\Vertex(5,50){2}
\Text(50,10)[b]{$X$}
\Text(50,90)[t]{$X$}
\end{picture}
\end{center}
with trivial defects inserted as shown.
By flattening the diagram above, as 
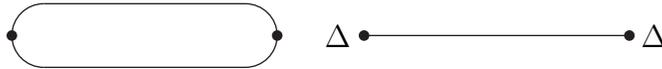
\begin{center}
\begin{picture}(120,50)
\CArc(22,24)(12,90,270)
\CArc(98,24)(12,-90,90)
\Line(22,36)(98,36)   \Line(22,12)(98,12)
\Vertex(10,24){2}  \Vertex(110,24){2}
\end{picture}
$\: \:$
\begin{picture}(120,50)
\Line(10,24)(110,24) 
\Vertex(10,24){2}  \Vertex(110,24){2}
\Text(5,24)[r]{$\Delta$}   
\Text(115,24)[l]{$\Delta$}
\end{picture}
\end{center}
and picking an orientation, say,
\begin{center}
\begin{picture}(120,50)
\ArrowLine(10,24)(110,24)
\Vertex(10,24){2}  \Vertex(110,24){2}
\Text(5,24)[r]{$\Delta$}   
\Text(115,24)[l]{$\Delta$}
\end{picture}
\end{center}
we predict that the closed string states are the same as elements of
\begin{displaymath}
{\rm Ext}^*_{X \times X}\left( 
\Delta^{\vee}, \Delta \right)
\end{displaymath}

As we shall discuss in section~\ref{hoch-closed}, this is the ``Hochschild
homology'' of $X$, and for smooth $X$, the Hochschild-Kostant-Rosenberg
(HKR) isomorphism \cite{andrei-utah}[theorem 6.3] says that
\begin{displaymath}
{\rm Ext}^*_{X \times X}\left( \Delta^{\vee}, \Delta \right)
\: = \:
\bigoplus_{p-q=*} H^p\left(X, \Omega^q_X \right)
\end{displaymath}
and so we see that the Ext groups on $X \times X$ are differential
forms, as expected for closed string states.  
Using the fact that
\begin{displaymath}
\Omega^q_X \: = \: K_X \otimes \Lambda^{n-q} TX
\end{displaymath}
where $n$ is the dimension of $X$, we have that
\begin{displaymath}
H^p\left(X, \Omega^q_X \right) \: = \: H^p\left(X, K_X \otimes
\Lambda^{n-q} TX \right)
\end{displaymath}
and so 
\begin{displaymath}
{\rm Ext}^*_{X \times X}\left( \Delta^{\vee}, \Delta \right)
\: = \:
\bigoplus_{p-q=*} H^p\left(X, \Omega^q_X \right)
\: = \:
\bigoplus_{p+q=n-*} H^p\left( X, K_X \otimes \Lambda^q TX \right)
\end{displaymath}

When $X$ is Calabi-Yau, the states above are well-known to match
closed string states on $X$ \cite{ed-tft}. 

The Hochschild-Kostant-Rosenberg (HKR) theorem above will form the intellectual
basis of the taffy identities for closed strings.  In effect, we will
use homological algebra to reduce all diagrams obtained by folding and
twisting closed strings to the diagram above, then apply HKR to argue
that the results match closed string states.

For our first example of a folded closed string, consider folding
a closed string into a U-bar shape, as
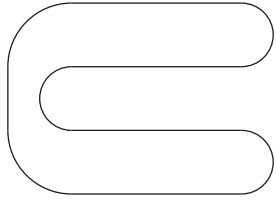
\begin{center}
\begin{picture}(100,80)
\Line(24,0)(88,0)  \Line(24,24)(88,24)
\Line(24,48)(88,48)  \Line(24,72)(88,72)
\CArc(88,12)(12,-90,90)  \CArc(88,60)(12,-90,90)
\CArc(24,36)(12,90,270)
\CArc(24,48)(24,90,180)  \CArc(24,24)(24,180,270)
\Line(0,48)(0,24)
\end{picture}
\end{center}
then we get a prediction that the closed string states are given by
\begin{equation}  \label{closed-U-1}
{\rm Ext}_{X \times X \times X \times X}^*\left(
\Delta_{14}^{\vee} \otimes \Delta_{23}^{\vee},
\Delta_{12} \otimes \Delta_{34}
\right).
\end{equation}
More generally, for ${\cal S} \in D^b(X^2 \times Y)$, it can be shown that
\begin{equation}   \label{closed-U-main}
{\rm Ext}^*_{X^4 \times Y} \left(
\Delta_{12}^{\vee}, \Delta_{14} \otimes \Delta_{23} \otimes
\pi_{34Y}^* {\cal S} \right)
\: = \:
{\rm Ext}^*_{X^2 \times Y} \left( \Delta_{12}^{\vee}, {\cal S} \right)
\end{equation}
which is our second ``taffy identity'' (and, for $Y$ a point and
${\cal S} = \Delta$, together with the HKR isomorphism,
implies that the states~(\ref{closed-U-1}) match closed
string states). 

The second taffy identity~(\ref{closed-U-main})
can be checked as follows \cite{andreipriv0}.
First, the left-hand-side can be written as the cohomology of
\begin{displaymath}
{\bf R} \Gamma\left( X^4 \times Y, \Delta_{12} \otimes \Delta_{23} \otimes
\Delta_{14} \otimes \pi_{34Y}^* {\cal S} \right)
\end{displaymath}
However, 
\begin{displaymath}
\Delta_{12} \otimes \Delta_{23} \otimes \Delta_{14} \: = \:
\Delta_{{\rm small}\, *} {\cal O}_{X \times Y}
\end{displaymath}
where
\begin{displaymath}
\Delta_{\rm small}: \: (x,y) \: \mapsto \: (x,x,x,x,y)
\end{displaymath}
(Intuitively, this $\Delta_{\rm small}$ is at least analogous to a small
loop between the various copies of $X$.)
Furthermore, 
\begin{displaymath}
\Delta_{ {\rm small}\, *} {\cal O}_{X \times Y} \otimes
\pi_{34Y}^* {\cal S} \: = \:
\Delta_{ {\rm small}\, *} \left( \Delta_{\rm small}^* \pi_{34Y}^*
{\cal S} \right)
\end{displaymath}
Thus, the left-hand-side of~(\ref{closed-U-main}) is the cohomology of
\begin{displaymath}
{\bf R} \Gamma\left( X \times Y, \Delta_{\rm small}^* \pi_{34Y}^* {\cal S} \right)
\end{displaymath}
Similarly, the right-hand-side of~(\ref{closed-U-main}) can be written
\begin{displaymath}
{\bf R} \Gamma\left( X^2 \times Y, \Delta_{12} \otimes {\cal S} \right)
\end{displaymath}
where $\Delta_{12} \equiv \Delta_{12 *} {\cal O}_{X \times Y}$ and
\begin{displaymath}
\Delta_{12}: \: (x,y) \: \mapsto \: (x,x,y)
\end{displaymath}
Now,
\begin{displaymath}
{\cal S} \otimes \Delta_{12 *} {\cal O}_{X \times Y}
\: = \:
\Delta_{12 *} \left( \Delta_{12}^* {\cal S} \right)
\end{displaymath}
so the right-hand-side of~(\ref{closed-U-main}) is
\begin{displaymath}
{\bf R} \Gamma\left( X \times Y, \Delta_{12}^* {\cal S} \right)
\end{displaymath}
Finally, the left- and right-hand-sides match because
$\pi_{34 Y} \circ \Delta_{\rm small} = \Delta_{12}$:
\begin{displaymath}
\xymatrix{
X \times Y \ar[r]^{\Delta_{\rm small} } 
\ar[dr]_{\Delta_{12} } & X^4 \times Y \ar[d]^{\pi_{34 Y} } \\
& X^2 \times Y
}
\end{displaymath}
Thus, we have established our second taffy identity.

Note that as a special case, if we take $Y$ to be the empty set,
and ${\cal S} = \Delta$, then identity~(\ref{closed-U-main}) reduces to
\begin{equation} \label{eq:hc-for-comp-to-hh}
{\rm Ext}^*_{X^4} \left( 
\Delta_{14}^{\vee} \otimes \Delta_{23}^{\vee},
\Delta_{12} \otimes \Delta_{34} \right)
\: = \:
{\rm Ext}^*_{X^2} \left(
\Delta^{\vee}, \Delta \right)
\end{equation}
as expected, which together with HKR verifies that~(\ref{closed-U-1})
describe closed string states.

For another application of the second taffy identity,
consider folding twice, to get
\begin{center}
\begin{picture}(100,120)
\Line(24,0)(88,0)  \Line(24,24)(88,24)
\Line(24,48)(76,48)  \Line(24,72)(76,72)
\Line(12,96)(76,96)  \Line(12,120)(76,120)
\CArc(88,12)(12,-90,90) 
\CArc(24,36)(12,90,270)
\CArc(24,48)(24,90,180)  \CArc(24,24)(24,180,270)
\Line(0,48)(0,24)
\CArc(12,108)(12,90,270)
\CArc(76,84)(12,-90,90)
\CArc(76,96)(24,0,90)  \CArc(76,72)(24,-90,0)
\Line(100,72)(100,96)
\end{picture}
\end{center}
and press down to form an open string on $X^6$,
then we get a prediction that the closed string states are given by
\begin{displaymath}
{\rm Ext}^*_{X^6}\left( \Delta_{14}^{\vee} \otimes \Delta_{23}^{\vee}
\otimes \Delta_{56}^{\vee}, 
\Delta_{12} \otimes \Delta_{36} \otimes \Delta_{45}
\right)
\end{displaymath}
We can show this by repeatedly applying identity~(\ref{closed-U-main})
to successively unfold the diagram above.
Applying it once yields
\begin{displaymath}
{\rm Ext}^*_{X^6}\left( \Delta_{14}^{\vee} \otimes \Delta_{23}^{\vee}
\otimes \Delta_{56}^{\vee}, 
\Delta_{12} \otimes \Delta_{36} \otimes \Delta_{45}
\right)
\: = \:
{\rm Ext}^*_{X^4}\left(
\Delta_{12}^{\vee},
\Delta_{34} \otimes \Delta_{23} \otimes \Delta_{14} \right)
\end{displaymath}
which we have already demonstrated to match closed string states.

It is straightforward to check that additional folds can be straightened;
we leave the details as an exercise for the reader.

In addition to folds, we can also stretch out sections.
For example, consider the diagram
\begin{center}
\begin{picture}(100,120)
\Line(24,0)(88,0)  \Line(24,24)(88,24)
\Line(24,48)(88,48)  \Line(24,72)(88,72)
\Line(24,96)(88,96)  \Line(24,120)(88,120)
\CArc(88,12)(12,-90,90)
\CArc(88,60)(12,-90,90)
\CArc(88,108)(12,-90,90)
\CArc(24,36)(12,90,270)   \CArc(24,84)(12,90,270)
\CArc(24,24)(24,180,270)   \CArc(24,96)(24,90,180)
\Line(0,24)(0,96)
\end{picture}
\end{center}
The string states in this case are
\begin{displaymath}
{\rm Ext}^*_{X^6}\left(
\Delta_{16}^{\vee} \otimes \Delta_{23}^{\vee} \otimes \Delta_{45}^{\vee},
\Delta_{12} \otimes \Delta_{34} \otimes \Delta_{56} \right)
\end{displaymath}
We can apply identity~(\ref{closed-U-main}) to show that the states above
match closed string states.  First, relabel the states above by exchanging
$4$ and $6$, to get
\begin{displaymath}
{\rm Ext}^*_{X^6}\left(
\Delta_{14}^{\vee} \otimes \Delta_{23}^{\vee} \otimes \Delta_{65}^{\vee},
\Delta_{12} \otimes \Delta_{36} \otimes \Delta_{54} \right)
\end{displaymath}
then, identity~(\ref{closed-U-main}) implies that this is the same as
\begin{displaymath}
{\rm Ext}^*_{X^4} \left(
\Delta_{12}^{\vee},
\Delta_{34} \otimes \Delta_{14} \otimes \Delta_{23} \right)
\end{displaymath}
which we have already shown matches closed string states.

Next, consider the three-pointed star:
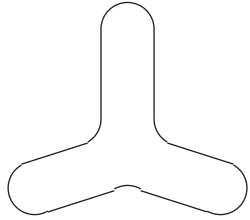
\begin{center}
\begin{picture}(100,100)
\CArc(50,90)(10,0,180)   
\CArc(85,30)(10,-120,60)  
\CArc(15,30)(10,120,300)   
\Line(40,90)(40,56)   \Line(60,90)(60,56)
\CArc(70,56)(10,180,240)   \CArc(30,56)(10,-60,0)
\Line(65,47)(90,39)
\Line(80,21)(55,29)
\Line(10,39)(35,47)
\Line(20,21)(45,29)
\CArc(50,21)(10,60,120)
\end{picture}
\end{center}
On the one hand, we can contract this to three open strings
joined at the center, with diagonal defects at each outer corner
and a more complicated diagonal defect at the center:
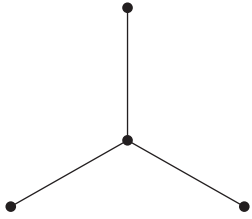
\begin{center}
\begin{picture}(100,100)
\Line(50,50)(50,100)  \Vertex(50,100){2}
\Line(50,50)(94,25)   \Vertex(94,25){2}
\Line(50,50)(6,25)    \Vertex(6,25){2}
\Vertex(50,50){2}
\end{picture}
\end{center}
On the other hand, we can `rolodex' this star into the fat E-shape figure
described previously, which gives us the states and also tells us that
they do match closed string states, as expected.

More generally, given an $n$-pointed star of the form above,
the states are given as
\begin{displaymath}
{\rm Ext}^*_{X^{2n}} \left(
\otimes_{i=1}^n \Delta_{2i,2i-1}^{\vee},
\otimes_{j=1}^n \Delta_{2j-1,2j-2} 
\right)
\end{displaymath}
and by repeatedly unravelling with identity~(\ref{closed-U-main}),
we can show that these match closed string states, as expected.

In addition to folds, we can also twist strings.
Consider first flattening a circle into a long flat oval,
then
twisting the oval into a figure eight.
To understand the spectrum,
fold at the twist to get a diagram of the form\footnote{
There is no information pertinent for these considerations contained
in whether lines cross over or under one another, hence we have not
tried to distinguish crossings in the picture shown.
}
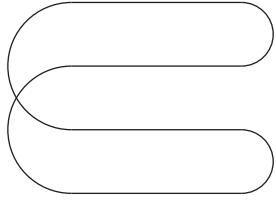
\begin{center}
\begin{picture}(100,80)
\Line(24,0)(88,0)  \Line(24,24)(88,24)
\Line(24,48)(88,48)  \Line(24,72)(88,72)
\CArc(88,12)(12,-90,90)  \CArc(88,60)(12,-90,90)
\CArc(24,24)(24,90,270)   \CArc(24,48)(24,90,270)
\end{picture}
\end{center}
then we get a prediction that the closed string states are given by
\begin{equation}  \label{closed-pretzel-1}
{\rm Ext}^*_{X^4} \left(
\Delta_{13}^{\vee} \otimes \Delta_{24}^{\vee},
\Delta_{12} \otimes \Delta_{34}  \right)
\end{equation}
which we will show momentarily to be correct.
More generally, it can be shown that \cite{andreipriv0} 
\begin{equation}  \label{closed-pretzel-main}
{\rm Ext}^*_{X^4 \times Y}\left(
\Delta_{12}^{\vee}, \Delta_{13} \otimes \Delta_{24} \otimes
\pi_{34Y}^* {\cal S} \right)
\: = \:
{\rm Ext}^*_{X^2 \times Y}\left(
\Delta_{12}^{\vee}, {\cal S} \right)
\end{equation}
for any ${\cal S} \in D^b(X^2 \times Y)$.
This is our third taffy identity.
This can be derived in almost exactly the same fashion as the second taffy
identity~(\ref{closed-U-main}); the only difference is that we
utilize the result that
\begin{displaymath}
\Delta_{12} \otimes \Delta_{13} \otimes \Delta_{24} \: = \:
\Delta_{\rm small}
\end{displaymath}
When $Y$ is the empty set and ${\cal S} = \Delta_{34}$,
this implies that
\begin{displaymath}
{\rm Ext}^*_{X^4}\left( \Delta_{12}^{\vee}, \Delta_{13} \otimes \Delta_{24}
\otimes \Delta_{34} \right) \: = \:
{\rm Ext}^*_{X^2}\left( \Delta^{\vee}, \Delta \right)
\end{displaymath}
which is precisely the conjecture~(\ref{closed-pretzel-1}).

For another example, 
consider taking a closed string, squeezing to a flattened
oval, then twisting twice and folding at each twist.
The resulting diagram is
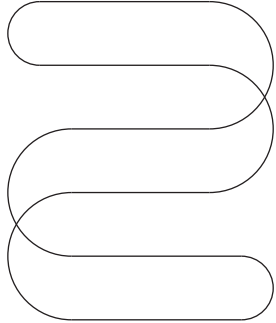
\begin{center}
\begin{picture}(100,120)
\Line(24,0)(88,0)  \Line(24,24)(88,24)
\Line(24,48)(76,48)  \Line(24,72)(76,72)
\Line(12,96)(76,96)  \Line(12,120)(76,120)
\CArc(88,12)(12,-90,90)
\CArc(12,108)(12,90,270)
\CArc(24,24)(24,90,270)  \CArc(24,48)(24,90,270)
\CArc(76,72)(24,-90,90)   \CArc(76,96)(24,-90,90)
\end{picture}
\end{center}
and the corresponding open string states are
\begin{displaymath}
{\rm Ext}^*_{X^6} \left(
\Delta_{13}^{\vee} \otimes \Delta_{24}^{\vee} \otimes \Delta_{56}^{\vee},
\Delta_{12} \otimes \Delta_{35} \otimes \Delta_{46} \right)
\end{displaymath}
We can repeatedly apply the identity~(\ref{closed-pretzel-main}) to 
successively unroll and untwist layers to show that
these are the same as closed string states on $X$.
After one application, one has that the states above are the same as
\begin{displaymath}
{\rm Ext}^*_{X^4}\left( \Delta_{12}^{\vee}, \Delta_{34} \otimes \Delta_{13}
\otimes \Delta_{24} \right)
\end{displaymath}
which is the case previously discussed, and so matches closed string states.

Because the taffy identities are insensitive to crossings, no knot invariants
can be built from these constructions.

\subsection{More general open-closed strings from defects}

For completeness, and later applications, let us make some easy general
observations on other more general examples of defects in the B model.

Consider, for our first example,
the case of
a closed string on $X$ with a single defect inserted:
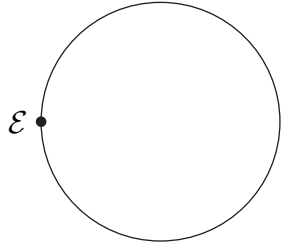
\begin{center}
\begin{picture}(100,100)(0,0)
\CArc(50,50)(45,0,360)
\Vertex(5,50){2}
\Text(0,50)[r]{${\cal E}$}
\end{picture}
\end{center}
defined by ${\cal E} \in D^b(X)$.
We can compute the open string states in this case by inserting
a trivial fold at the opposite side of the circle:
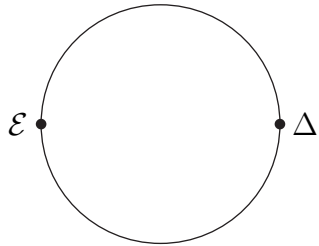
\begin{center}
\begin{picture}(100,100)(0,0)
\CArc(50,50)(45,0,360)
\Vertex(5,50){2}
\Vertex(95,50){2}
\Text(0,50)[r]{${\cal E}$}
\Text(100,50)[l]{$\Delta$}
\end{picture}
\end{center}
then from the previous analysis, we see that the open string specturm
is given by
\begin{displaymath}
{\rm Ext}^*_{X \times X}\left( \Delta_* {\cal E}, \Delta
\right)
\end{displaymath}

Next, let us generalize to ``closed strings'' formed from gluing together
open strings on distinct spaces along defects.
To begin, consider
an infinite cylinder split lengthwise into a pair of
semi-infinite strips joined along two edges, one string on $X$ and the
other on $Y$, so that a spacelike cross-section is
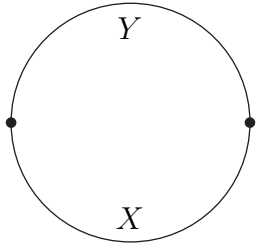
\begin{center}
\begin{picture}(100,100)(0,0)
\CArc(50,50)(45,0,180)
\CArc(50,50)(45,180,360)
\Vertex(95,50){2}
\Vertex(5,50){2}
\Text(50,10)[b]{$X$}
\Text(50,90)[t]{$Y$}
\end{picture}
\end{center}
with defects defined by ${\cal S}_1, {\cal S}_2 \in D^b(X \times Y)$.
This is physically equivalent to an ordinary open string on $X \times Y$:
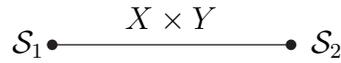
\begin{center}
\begin{picture}(120,20)(0,0)
\Line(15,10)(105,10)
\Vertex(105,10){2}
\Vertex(15,10){2}
\Text(60,15)[b]{$X \times Y$}
\Text(12,10)[r]{${\cal S}_1$}
\Text(113,10)[l]{${\cal S}_2$}
\end{picture}
\end{center}
In particular, in both cases the open string states (inserted in the
infinite future or past) are
\begin{equation}   \label{split-string}
{\rm Ext}^*_{X \times Y}\left( {\cal S}_1, {\cal S}_2 \right)
\end{equation}
It is somewhat tempting to incorrectly speculate that the
open string states in the first picture arise from both
\begin{equation} \label{bad-split-string}
{\rm Ext}^*_X\left( \pi_{1 *} {\cal S}_1, \pi_{1 *} {\cal S}_2 \right),
\: \: \:
{\rm Ext}^*_Y\left( \pi_{2 *} {\cal S}_1, \pi_{2 *} {\cal S}_2 \right)
\end{equation}
however, this is incorrect, as the corresponding operators are inserted
at points where $X$ and $Y$ come together, so one cannot meaningfully
distinguish in the form implied in~(\ref{bad-split-string}).
(Moreover, there is no mathematical equivalence between
(\ref{split-string}) and (\ref{bad-split-string}).)

With larger numbers of segments, additional folding tricks are possible.
Suppose there are four segments, so that a cross-section of the
partitioned cylinder is
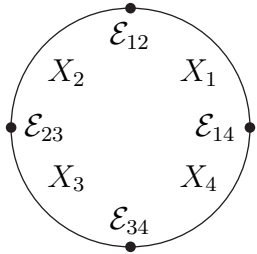
\begin{center}
\begin{picture}(100,100)(0,0)
\CArc(50,50)(45,0,90)   \CArc(50,50)(45,90,180)
\CArc(50,50)(45,180,270)  \CArc(50,50)(45,270,360)
\Vertex(95,50){2}  \Vertex(50,95){2}
\Vertex(5,50){2}  \Vertex(50,5){2}
\Text(76,76)[t]{$X_1$}   \Text(26,76)[t]{$X_2$}
\Text(26,26)[b]{$X_3$}  \Text(76,26)[b]{$X_4$}
\Text(90,50)[r]{${\cal E}_{14}$}
\Text(50,90)[t]{${\cal E}_{12}$}
\Text(10,50)[l]{${\cal E}_{23}$}
\Text(50,10)[b]{${\cal E}_{34}$}
\end{picture}
\end{center}
where ${\cal E}_{ij} \in D^b(X_i \times X_j)$,
which folds into the equivalent diagram
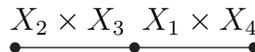
\begin{center}
\begin{picture}(100,50)(0,0)
\Line(95,25)(50,25)
\Line(50,25)(5,25)
\Vertex(50,25){2}
\Text(25,30)[b]{$X_2 \times X_3$}
\Text(75,30)[b]{$X_1 \times X_4$}
\Vertex(95,25){2}  \Vertex(5,25){2}
\end{picture}
\end{center}
with middle defect on
\begin{displaymath}
\pi_{12}^* {\cal E}_{12} \otimes
\pi_{34}^* {\cal E}_{34}
\end{displaymath}
which folds again into an open string on 
\begin{displaymath}
Y \: \equiv \: X_1 \times X_2 \times X_3 \times X_4
\end{displaymath}
with open string states
\begin{displaymath}
{\rm Ext}_Y^*\left( \pi_{12}^* {\cal E}_{12} \otimes
\pi_{34}^* {\cal E}_{34},
\pi_{23}^* {\cal E}_{23}^{\vee} \otimes
\pi_{14}^* {\cal E}_{14}
\right)
\end{displaymath}

Another folding trick involves a cylinder formed from three open strings:
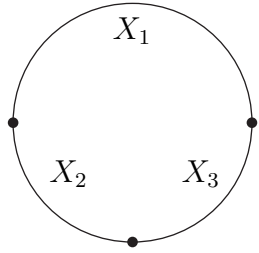
\begin{center}
\begin{picture}(100,100)(0,0)
\CArc(50,50)(45,0,180)   
\CArc(50,50)(45,180,270)  \CArc(50,50)(45,270,360)
\Vertex(95,50){2}  
\Vertex(5,50){2}  \Vertex(50,5){2}
\Text(50,90)[t]{$X_1$}
\Text(26,26)[b]{$X_2$}  \Text(76,26)[b]{$X_3$}
\end{picture}
\end{center}
We first split the top open string by inserting $\Delta_* {\cal O}_{X_1}$, as
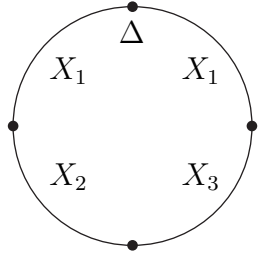
\begin{center}
\begin{picture}(100,100)(0,0)
\CArc(50,50)(45,0,90)   \CArc(50,50)(45,90,180)
\CArc(50,50)(45,180,270)  \CArc(50,50)(45,270,360)
\Vertex(95,50){2}  \Vertex(50,95){2}
\Vertex(5,50){2}  \Vertex(50,5){2}
\Text(76,76)[t]{$X_1$}   \Text(26,76)[t]{$X_1$}
\Text(26,26)[b]{$X_2$}  \Text(76,26)[b]{$X_3$}
\Text(50,90)[t]{$\Delta$}
\end{picture}
\end{center}
and then fold it into 
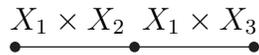
\begin{center}
\begin{picture}(100,50)(0,0)
\Line(95,25)(50,25)
\Line(50,25)(5,25)
\Vertex(50,25){2}  \Vertex(95,25){2}  \Vertex(5,25){2}
\Text(25,30)[b]{$X_1 \times X_2$}
\Text(75,30)[b]{$X_1 \times X_3$}
\end{picture}
\end{center}
which, after another fold, has states counted by
\begin{displaymath}
{\rm Ext}^*_{Y'}\left(\pi_{12}^* {\cal E}_{12} \otimes
\pi_{31}^* {\cal E}_{31}^{\vee}, \pi_{23}^* {\cal E}_{23} \otimes
\pi_{11}^* \Delta_* {\cal O}_{X_1} \right)
\end{displaymath}
where
\begin{displaymath}
Y' \: \equiv \: X_1 \times X_1 \times X_2 \times X_3
\end{displaymath}

Note that we can efficiently specify locations of parallel defects along
a cylinder through a simplex:  specify the location of each defect 
by what fraction of the circumference it sits at, relative to the previous
defect.  If there are $k$ defects, then that gives us $k$ real numbers
between $0$ and $1$ whose sum is necessarily $1$, which is precisely
a simplex.

Another folding trick involves a diagram that cannot be understood
as just a cylinder, with cross section
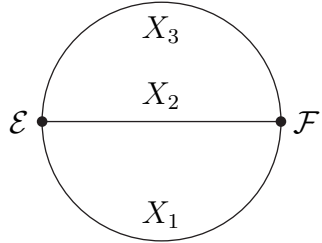
\begin{center}
\begin{picture}(100,100)(0,0)
\CArc(50,50)(45,0,180)
\CArc(50,50)(45,180,360)
\Line(95,50)(5,50)
\Vertex(95,50){2}
\Vertex(5,50){2}
\Text(50,10)[b]{$X_1$}
\Text(50,55)[b]{$X_2$}
\Text(50,90)[t]{$X_3$}
\Text(100,50)[l]{${\cal F}$}
\Text(0,50)[r]{${\cal E}$}
\end{picture}
\end{center}
where ${\cal E}, {\cal F} \in D^b(X_1 \times X_2 \times X_3)$.
This folds into the equivalent open string on $X_1 \times X_2 \times X_3$:
\begin{center}
\begin{picture}(100,50)(0,0)
\Line(5,25)(95,25)
\Vertex(95,25){2}
\Vertex(5,25){2}
\Text(50,30)[b]{$X_1 \times X_2 \times X_3$}
\Text(100,25)[l]{${\cal F}$}
\Text(0,25)[r]{${\cal E}$}
\end{picture}
\end{center}
from which we see that the string states arising from this diagram are
\begin{displaymath}
{\rm Ext}^*_{X_1 \times X_2 \times X_3}\left(
{\cal E}, {\cal F} \right)
\end{displaymath}

\subsection{Correlation functions}

So far we have only checked that massless states match after
performing taffy operations.  In this section, we shall outline
how correlation function matching should occur.  Unlike the
case of massless states, for correlation functions we do not have
rigorous proofs, so we will only outline conjectures.

Let us begin by considering correlation functions on a disk,
describing an open string on $X$,
with the same boundary conditions everywhere along the edge of the
disk, corresponding to a object ${\cal E}$ in $D^b(X)$.
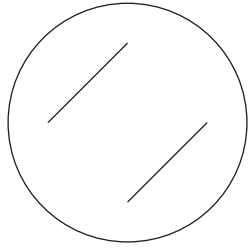
\begin{center}
\begin{picture}(100,100)
\CArc(50,50)(45,0,360)
\Line(20,50)(50,80)
\Line(50,20)(80,50)
\end{picture}
\end{center}
Now, split the disk in half along the identity defect, as
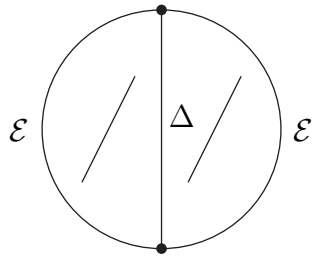
\begin{center}
\begin{picture}(100,100)
\CArc(50,50)(45,0,360)
\Line(50,5)(50,95)
\Vertex(50,5){2}  \Vertex(50,95){2}
\Text(0,50)[r]{${\cal E}$}  \Text(100,50)[l]{${\cal E}$}
\Text(53,55)[l]{$\Delta$}
\Line(20,30)(40,70)
\Line(60,30)(80,70)
\end{picture}
\end{center}
To perform such a splitting, we need to insert boundary-condition-changing
operators at top and bottom intersection
points, which in the present case will be
elements of  
\begin{equation}  \label{bound-op-def-insert}
{\rm Ext}^*_{X \times X}\left( \pi_1^* {\cal E}^{\vee} \otimes
\pi_2^* {\cal E}, \Delta \right)
\end{equation}
This group is the same as
\begin{equation}  \label{bound-op-def-insert-2}
{\rm Ext}^*_X\left( {\cal E}, {\cal E} \right)
\end{equation}
which matches the intuition that inserting the diagonal defect is
equivalent to doing nothing, since if we did not insert the defect,
the boundary operators would have been counted by~(\ref{bound-op-def-insert-2}).

We can then fold the diagram above, to obtain the disk diagram
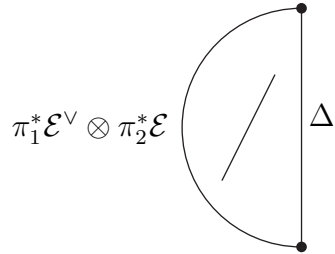
\begin{center}
\begin{picture}(100,100)
\CArc(50,50)(45,90,270)
\Line(50,5)(50,95)
\Vertex(50,5){2}  \Vertex(50,95){2}
\Text(0,50)[r]{$\pi_1^* {\cal E}^{\vee} \otimes
\pi_2^* {\cal E}$}
\Text(53,55)[l]{$\Delta$}
\Line(20,30)(40,70)
\end{picture}
\end{center}
on $X \times X$.

In principle, we expect correlation functions on the folded disk
to match correlation
functions on the original disk.

{\it Conjecture:}  The open string theory formed by taking a closed
string worldsheet and triangulating into open strings with defect
boundaries, is equivalent to the original closed string theory.

This would be in analogy with the behavior of two-dimensional
QCD (see for example \cite{cmr}).
In the two-dimensional QCD story, one triangulates a Riemann surface, roughly,
by inserting traces over group representations along edges.
Here, by contrast, one is not inserting a complete set of states,
but rather is merely inserting a trivial defect,
creating a trivial reparametrization of the worldsheet, no more.

The conjecture above implies that one could fold all closed string
diagrams into open string disk diagrams.  For example, a sphere on $X$
could
be flattened to a disk on $X \times X$ with the diagonal $\Delta$
along the edge:
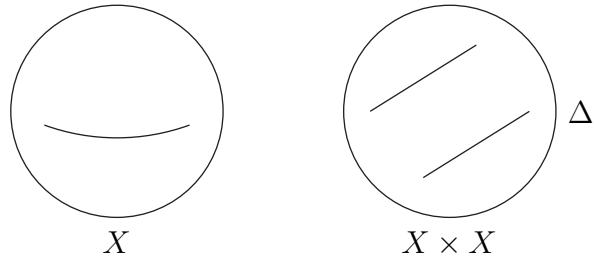
\begin{center}
\begin{picture}(100,100)
\CArc(50,50)(40,0,360)
\CArc(50,120)(80,250,290)
\Text(50,5)[t]{$X$}
\end{picture}
$\: \: \:$
\begin{picture}(120,100)
\CArc(60,50)(40,0,360)
\Line(30,50)(70,75)
\Line(50,25)(90,50)
\Text(105,50)[l]{$\Delta$}
\Text(60,5)[t]{$X \times X$}
\end{picture}
\end{center}
Similarly, a two-torus on $X$ can be flattened into an annulus on $X \times X$:
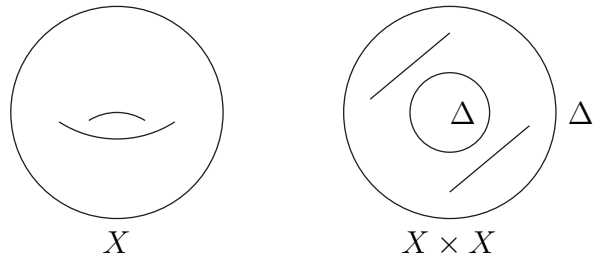
\begin{center}
\begin{picture}(100,100)
\CArc(50,50)(40,0,360)
\CArc(50,80)(40,237,303)
\CArc(50,30)(20,58,122)
\Text(50,5)[t]{$X$}
\end{picture}
$\: \: \:$
\begin{picture}(120,100)
\CArc(60,50)(40,0,360)
\CArc(60,50)(15,0,360)
\Line(30,55)(60,80)
\Line(60,20)(90,45)
\Text(105,50)[l]{$\Delta$}
\Text(70,50)[r]{$\Delta$}
\Text(60,5)[t]{$X \times X$}
\end{picture}
\end{center}
which can then be folded into a disk on $X^4$:
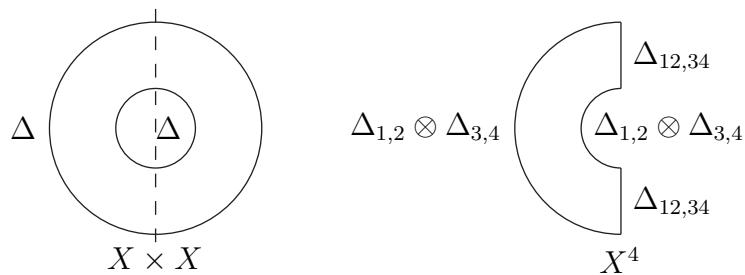
\begin{center}
\begin{picture}(120,100)
\CArc(60,50)(40,0,360)
\CArc(60,50)(15,0,360)
\DashLine(60,95)(60,7){5}
\Text(15,50)[r]{$\Delta$}
\Text(70,50)[r]{$\Delta$}
\Text(60,5)[t]{$X \times X$}
\end{picture}
$\: \: \:$
\begin{picture}(140,100)
\CArc(100,50)(40,90,270)
\CArc(100,50)(15,90,270)
\Line(100,90)(100,65)
\Line(100,10)(100,35)
\Text(55,50)[r]{$\Delta_{1,2} \otimes \Delta_{3,4}$}
\Text(90,50)[l]{$\Delta_{1,2} \otimes \Delta_{3,4}$}
\Text(105,77)[l]{$\Delta_{12,34}$}
\Text(105,22)[l]{$\Delta_{12,34}$}
\Text(100,5)[t]{$X^4$}
\end{picture}
\end{center}
Similar manipulations can be performed at higher genera, reducing all
such diagrams to disk diagrams on products of copies of $X$.

\section{String topology}
\label{string-top}

The taffy identities in \S\ref{bmodel:string-states} arose from
studying the homological algebra of the category of chain complexes
of $\O_{X}$-modules, where $X$ was a Calabi-Yau manifold.  The
ingredients which lead to these identities are available in other
contexts.  In this section we shall outline the details for the
case of `string topology.'  This can be described as
a mathematical abstraction of
bosonic string field theory (see {\it e.g.} \cite{sullivan1,sullivan2}),
that is well-known in the homotopy
community.  
In this section we shall derive precise analogues of the taffy
identities for string topology.

Let $X$ be a manifold, and let $LX$ be its free loop space.  For simplicity,
we will assume in this section that $X$ is simply-connected
\cite{goodwillie}.  (See \cite{lurietft}[remark 4.2.17] for an outline
of the non-simply-connected case.)  
Costello
(\cite{costello}; see also \cite[section 4.2]{lurietft}) explains how
the string topology operations of Chas and Sullivan on $C^{*}LX$, the
rational cochains on the free loop space,  can be understood as
arising from an open-closed TFT on $C^{*}X.$  
In this language, the closed string states are\footnote{
We have not constructed vertex operators to physically realize such in a CFT;
rather, we are saying that formally, the object playing the role of
closed string states is $H^*(LX)$.  A similar statement is true for open
string states here.
} $H^*(LX)$.

Let $C^{*}X$ be its algebra of rational singular
cochains, which plays the role of the open string algebra.   
There is an $\infty$-category $M_{X}$ of $C^{*}X$-modules,
which plays the role for $C^{*}X$ analogous to the $\infty$-category $D_{b}^{\infty} (X)$ of
quasi-coherent sheaves, whose homotopy category is the derived
category (see Remark \ref{rem-1}).   
In other words, (complexes of)
D-branes are elements of $M_X$, $C^*X$ modules.
In this language, open string states between two (complexes of)
D-branes
${\cal E}, {\cal F} \in M_X$ are given by
\begin{displaymath}
{\bf R}{\rm Hom}\left( {\cal E}, {\cal F} \right)
\end{displaymath}
(Unlike algebraic geometry, here there is no distinction between local and
global Hom; working over rational cochains is more closely analogous
to working on an affine scheme, where the module defining the sheaf is
the same as the global sections.  Therefore, we use the same Hom to
describe both $M_X$ modules and also the derived functor of global sections.)

Now, we shall start working out taffy identities.
If $f: X \to Y$ is a map of spaces, then associated to
the pull-back of singular cochains
\[
     f^{*}:\:  C^{*}Y \: \longrightarrow \: C^{*}X
\]
we have the derived pushforward
\[
     f_{*}: \: M_{X} \: \longrightarrow \:  M_{Y},
\]
which has a left adjoint 
\[
    f^{*}:\: M_{Y} \: \longrightarrow \: M_{X}.
\]
If $\cE$ and $\cF$ are $C^{*}X$-modules, then we write ${\bf R}
\hom_{X} (\cE,\cF)$ for
${\bf R} \hom_{M_{X}} (\cE,\cF)$; it is again a $C^{*}X$-module.  
Moreover, we can
define $\dual{E} = {\bf R}\hom_{X} (\cE,C^{*}X)$.  Provided that $\cE$ is in a
suitable sense finitely generated over $C^{*}X$ (we will use the term
``dualizable''), we have equivalences 
\begin{align*}
     \dual{(\dual{\cE})}        & \: \heq \: \cE \\
    \dual{\cE}\otimes \cF & \: \heq \: {\bf R}\hom_{X} (\cE,\cF).
\end{align*}

Let $\cE$ and $\cF$ be dualizable $C^{*}X$-modules, and let $\Delta:
X\to X\times X$ 
be the diagonal.  Then 
\begin{align*}
   {\bf R} \hom_{X\times X} (\pi_1^* \cE \otimes \pi_2^* \dual{\cF}, \Delta_{*}C^*{X})
\: &\heq \: {\bf R} \hom_{X} (\Delta^{*} (\pi_1^* \cE \otimes \pi_2^* \dual{\cF}), C^*{X}) \\
   &\heq \: {\bf R} \hom_{X} (\cE\otimes \dual{\cF},C^*{X}) \\
   & \heq \: {\bf R} \hom_{X} (C^*{X},\dual{\cE}\otimes \cF) \\
   &  \heq \: {\bf R}\hom_{X} (\cE,\cF),
\end{align*}
which is the analogue of the taffy identity \eqref{fold-strip-U}.

The analogue the projection formula \eqref{eq:2} 
holds in this context.  
If $f: X\to Y$, $\cE$ is a $C^{*}X$-module, and $\cG$ is 
$C^{*}Y$-module, then 
\[
    f_{*} (\cE \otimes_{X} f^{*}\cG)\: \heq \: ( f_{*}\cE )\otimes_{Y}\cG,
\]
and, at least for products, the Eilenberg-Zilber Theorem provides the
analog of the flat base change theorem (Prop 6).  These considerations
immediately yield the taffy identity analogous to \eqref{U-shape-main}
\[
{\bf R} \hom_{X\times X\times Y} (\pi_{1}^{*}E \otimes \pi_{23}^{*}\dual{S},
\pi_{12}^{*} \Delta) \: \heq \: {\bf R} \hom_{X\times Y} (\pi_{1}^{*}E, S)
\]
for any dualizable $C^{*} (X\times Y)$-module $S$.

Similarly, if $S$ is a  $C^{*} (X^{2}\times Y)$-module, and if
$\Delta$ refers to the $C^{*} (X^{2}\times Y)$-module 
\[
   \Delta = \Delta_{*} (C^{*} (X\times Y))
\]
obtained by pushing forward along $(x,y)\mapsto (x,x,y)$,
and if
\[
   \Delta_{ij} = \pi_{ij}^{*}\Delta
\]
with $\pi_{ij}$ projection to the indicated factors 
\[
   \pi_{ij}: X^{4} \times Y \to X^{2} \times Y,
\]
then the taffy identities \eqref{closed-U-main} and
\eqref{closed-pretzel-main} become
\begin{align*}
    \Rhom_{X^{4}\times Y} 
(\Delta_{12}^{\vee}, \Delta_{14}\otimes \Delta_{23}\otimes
\pi_{34}^{*}S) & \heq  \Rhom_{X^{2}\times Y} (\Delta^{\vee}, S) \\
    \Rhom_{X^{4}\times Y} 
(\Delta_{12}^{\vee}, \Delta_{13}\otimes \Delta_{24}\otimes
\pi_{34}^{*}S) & \heq  \Rhom_{X^{2}\times Y} (\Delta^{\vee}, S). \\
\end{align*}
The proofs follow the same pattern as the proof we gave of
\eqref{closed-U-main}, using the projection formula and the flat base
change theorem already mentioned. 

A more compelling observation is that the 
the closed string taffy identity  
associated to the flattening 
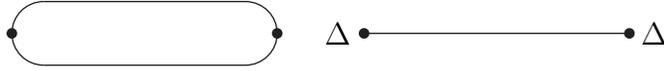
\begin{center}
\begin{picture}(120,50)
\CArc(22,24)(12,90,270)
\CArc(98,24)(12,-90,90)
\Line(22,36)(98,36)   \Line(22,12)(98,12)
\Vertex(10,24){2}  \Vertex(110,24){2}
\end{picture}
$\: \:$
\begin{picture}(120,50)
\Line(10,24)(110,24) 
\Vertex(10,24){2}  \Vertex(110,24){2}
\Text(5,24)[r]{$\Delta$}   
\Text(115,24)[l]{$\Delta$}
\end{picture}
\end{center}
predicts that the closed string algebra associated to $C^{*}X$ should
be 
\[
    {\bf R} \hom_{X \times X} (\dual{\Delta},\Delta) \heq
    \Delta\otimes^{\mathbf{L}}_{C^{*} (X\times  X)} \Delta.
\]
As we shall outline below, it is
well-known that this chain complex is equivalent both to the
Hochschild chains on $C^{*}X$, $HC_{*} (C^{*}X; C^{*}X)$, and to the
cochains on the free loop space $LX=\Map (S^{1},X)$, $C^{*} (LX)$.
This will give us the string-topology-analogue of the
Hochschild-Kostant-Rosenberg isomorphism, that played a crucial role
in the B model.

Let us consider this situation mathematically.
It turns out that the mathematics of the situation is quite close to the
defects we discuss here.  By introducing two points (defects) on the
closed string word sheet, we can identify the space of closed strings
(``loops'') in $X$  with the space of pairs of paths in $X$, joined at
their endpoints.  That is, the loop space $LX = {\rm Map} (S^{1},X)$ is the
pull-back of the diagram
 \begin{equation}\label{eq:st2}
\xymatrix{
& &
{PX}
\ar[d]^{ev_{01}} 
\\
{PX}
\ar[rr]^{ev_{01}}
& &
{X\times X.}
}
\end{equation}
Here $PX = {\rm Map} ([0,1],X)$ is the space of maps of the unit interval
to $X$, and $ev_{ij}: PX \to X^{2}$ is the map obtained by evaluating
the $0$-endpoint into the $i$ factor, and the $1$-endpoint into the
$j$ factor in the product.

Homotopically, it is equivalent to shrink the paths to constant length,
yielding the diagram
\begin{equation}\label{eq:st1}
\xymatrix{
& &
{X}
\ar[d]^{\Delta}
\\
{X}
\ar[rr]^-{\Delta}
& &
{X\times X.}
}
\end{equation}
The point-set pull-back in the diagram \eqref{eq:st1} is not $LX$,
but merely $X$.  Since homotopy functors, such as
cohomology, $K$-theory, and the derived category, view these diagrams
as equivalent, they cannot possibly preserve pull-backs.  It is
necessary to consider in this situation a homotopy-invariant
``derived pull-back''.

It turns out that one can construct an object which encodes all
possible homotopy deformations of the diagram \eqref{eq:st1}: it is the
``cosimplicial space''  $B^{\bullet} (X,X\times X, X)$,
\[
\xymatrix{
{\ldots} \\
{X\times Y^{2} \times X}
\ar@<-6ex>[u] \ar@<-2ex>[u]   \ar@<+2ex>[u] \ar@<+6ex>[u]
\\
{X\times Y \times X}
 \ar@<-3ex>[u]  \ar[u] \ar@<+3ex>[u]\\
{X\times X}
 \ar@<-2ex>[u] \ar@<+2ex>[u]
}
\]
Here we have written $Y$ for $X\times X$.  The maps are
\begin{align*}\label{}
   d^{0} (x_{0},y_{1},\dotsc ,y_{n-1},x_{n}) & = 
        (x_{0},\Delta (x_{0}),y_{1},\dotsc ) \\
   d^{n} (x_{0},y_{1},\dotsc ,y_{n-1},x_{n}) & = 
         (x_{0},y_{1},\dotsc ,y_{n-1},\Delta (x_{n}), x_{n})\\
   d^{i} (x_{0},y_{1},\dotsc ,y_{n-1},x_{n}) &= 
         (x_{0},y_{1},\dotsc ,y_{i}, y_{i},\dotsc ,y_{n-1},x_{n}) &&
         {1\leq i\leq n-1}.
\end{align*}

A basic theorem in homotopy theory
(\cite{Adams:Cobar},\cite[esp. p. 268]{BousfieldKan},\cite{Rector:EM})
implies that the situation gets no 
more complicated than the pull-back diagram \eqref{eq:st2}:

\begin{Proposition}\label{t-pr-hochschild-loop}
If $X$ is simply connected, then
\[
    B^{\bullet} (X,X\times X, X) \heq LX.
\]
\end{Proposition}

By applying a homotopy functor which preserves derived pull-backs, we
get more familiar results.  For
example, applying singular cochains $C^{*} (-)$, we find that
\begin{equation}\label{eq:st3}
      B_{\bullet}(C^{*}X,C^{*} (X\times X),C^{*}X) \heq C^{*} (LX).
\end{equation}
The left-hand side in \eqref{eq:st3} refers to the bar
complex which calculates ${\rm Tor}$, and so taking cohomology on both
sides yields
\[
   {\rm Tor}^{C^{*} (X\times X)} (C^{*}\Delta,C^{*}\Delta) = H^{*} (LX).
\]
Using the relation between Tor and Ext groups, we can rewrite this as
\begin{displaymath}
{\bf R}{\rm Hom}_{X \times X}\left( C^* \Delta^{\vee}, C^* \Delta \right)
\: = \: H^*(LX)
\end{displaymath}
Moreover, it is a theorem of Cartan and Eilenberg \cite{cart-eil}[section IX.6]
that the
chain complex associated to $B_{\bullet} (C^{*}X,C^{*} (X\times X),
C^{*}X)$ is equivalent to the cyclic complex which calculates
Hochschild homology:
\[
   HC_{*} (C^{*}X) \heq B_{\bullet} (C^{*}\Delta,C^{*} (X\times X), C^{*}\Delta),
\]
and so we have the following (well-known, see for example
\cite{Cohen:CyclicLectureNotes}[theorem 1.5.1], \cite{jones,gjp}) result.

\begin{Proposition}  \label{string-top-hkr}
If $X$ is simply connected, then
\[
   HH_{*} (C^{*}X) \: = \: H^{*} (LX) \: = \: \Tor^{C^{*} (X\times X)}
   (C^{*}\Delta,C^{*}\Delta)
\: = \: {\bf R}{\rm Hom}_{X \times X}\left(
C^* \Delta^{\vee}, C^* \Delta \right). 
\]
\end{Proposition}
This is precisely the analogue of the Hochschild-Kostant-Rosenberg isomorphism,
expressing closed string states on $X$ in terms of open string states
on $X \times X$, and in terms of the Hochschild homology of 
the open string algebra $C^* X$.

Hopkins and Lurie (\cite{lurietft,hopkins-lurie}; see also Blumberg,
Cohen, and Teleman \cite{bct-octft}) have shown that $C^{*}X$ is the
open string algebra of an open-closed TFT, whose closed string algebra
is $C^{*} (LX);$ and they have shown that the  resulting   structure
on $C^{*} (LX)$ includes the string topology operations of Chas and
Sullivan.

We emphasize that the identification of $LX$ as the derived-pull back
of
\[
   X \xrightarrow{\Delta} X\times X \xleftarrow{\Delta} X
\]
encodes the closed string taffy identities we consider in
\S\ref{bmodel:string-states}, at the level of topological spaces: they
then have instances in many algebraic settings by applying homotopy
functors.  To give an example,  consider a loop in $X$ viewed as
assembled from four paths, as in
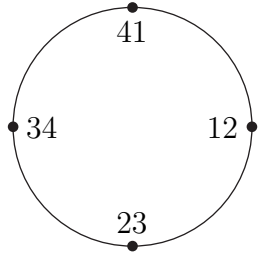
\begin{center}
\begin{picture}(100,100)(0,0)
\CArc(50,50)(45,0,90)   \CArc(50,50)(45,90,180)
\CArc(50,50)(45,180,270)  \CArc(50,50)(45,270,360)
\Vertex(95,50){2}  \Vertex(50,95){2}
\Vertex(5,50){2}  \Vertex(50,5){2}
\Text(90,50)[r]{$12$}
\Text(50,90)[t]{$41$}
\Text(10,50)[l]{$34$}
\Text(50,10)[b]{$23$}
\end{picture}
\end{center}
That is, the loop space is the pull-back in the diagram
\[
\xymatrix{
& &
{PX \times PX}
\ar[d]^{ev_{12}\times ev_{34}} 
\\
{PX \times PX}
\ar[rr]^{ev_{14}\times ev_{23}}
& &
{X^{4}.}
}
\]
But this diagram is homotopy equivalent to
\[
\xymatrix{
& &
{X \times X}
\ar[d]^{\Delta_{12}\times \Delta_{34}} 
\\
{X\times X}
\ar[rr]^{\Delta_{14}\times \Delta_{23}}
& &
{X^{4}.}
}
\]
The same analysis as Proposition \ref{t-pr-hochschild-loop} shows
that, if $X$ is simply connected, then
\[
    B^{\bullet} (\Delta_{12}\times \Delta_{34}, X^{4},
    \Delta_{14}\times \Delta_{23}) \heq LX, 
\]
and so by applying singular cochains we find that
\[
    \Tor^{C^{*} (X^{4})} (C^{*} (\Delta_{12})\otimes C^{*} (\Delta_{34}), C^{*}
    ( \Delta_{14})\otimes C^{*} (\Delta_{23})) = \Tor^{C^{*} (X^{2})} (C^{*}
    (\Delta),C^{*} (\Delta)) = H^{*} (LX),
\]
with all of these agreeing with the Hochschild homology of the
signular cochains, $HH_{*} (C^{*}X)$.  This is the analogue, for
string topology, of the taffy identity \eqref{eq:hc-for-comp-to-hh}
for closed strings in the $B$-model.

\section{Hochschild (co)homology and closed string states}
\label{hoch-closed}

It is sometimes said that closed string states are the Hochschild (co)homology
of the open string algebra.  In the physics literature, closed string
states are often related to Hochschild cohomology; 
see for example \cite{kr}[section 2],
\cite{ms}[section 3.3] and
references therein.
In the mathematics community, closed string states are often related to
Hochschild homology\footnote{
M.A. first learned that closed string states are related to open string
states in this way from D.~Berenstein in 2001.
}, see work of {\it e.g.}
Kontsevich, Costello, Hopkins--Lurie\footnote{
The
work of Hopkins and Lurie places this relationship in the context of
enriched topological field theory, which we outlined in
section~\ref{higher-cat}.
}, Blumberg--Cohen--Teleman \cite{lurietft,costello,hopkins-lurie,bct-octft}.
In this section we shall examine this correspondence.
Among other things, we shall argue that, based on the taffy identities,
the natural relation is between
the closed string states and Hochschild homology instead of cohomology.

It is also worth pointing out specifically that
Costello, Hopkins, and Lurie
\cite{lurietft,costello,hopkins-lurie}
have developed an approach to topological field theory
in which the rotational symmetry of the closed string states is reflected
in Connes' cyclic structure on the Hochschild complex, and so cyclic homology
also naturally appears in their framework (see {\it e.g.}
\cite{lurietft}[section 4.2]).  In physics, the relationship between cyclic
homology and closed (bosonic) strings was recently discussed in
\cite{moeller-sachs}.  (As string topology is a mathematical extraction of
bosonic string field theory, there are close parallels between their work
and parts of our discussion of string topology; our 
proposition~\ref{string-top-hkr}, for example,
is essentially \cite{moeller-sachs}[equ'n~(95)].)

\subsection{B model}

In the language of the B model topological field theory,
closed string states can be related to Hochschild (co)homology
via 
the ``Hochschild-Kostant-Rosenberg (HKR) isomorphism,''
\begin{displaymath}
H^*\left(X, \Lambda^* TX \right) \: = \:
{\rm Ext}^*_{X \times X}\left( \Delta, \Delta \right)
\end{displaymath}
for a Calabi-Yau $X$.
One, somewhat vague, reason sometimes stated for this relationship
between closed and open strings is that
if one accepts that closed strings should be derived from open strings,
then closed string states should arise as some sort of cohomology operation
on the open string algebra, and Hochschild cohomology arises very naturally
in this role.  A more refined intuition is sometimes stated, in terms of
folding closed strings into open strings.  

In section~\ref{bmodel-closed} we learned how to make such intuitions
physically precise for the B model.  
In the process, we learned that the closed string
states are naturally given by Hochschild homology, not cohomology.
In this section we will review Hochschild homology and cohomology, and
their relationship.

In general, for an algebra $A$, the Hochschild homology is 
\cite{loday}[prop. 1.1.13]
\begin{displaymath}
HH_*(A) \: = \: {\rm Tor}^{A \otimes A^{op}}_* \left( A, A \right)
\end{displaymath}
and the Hochschild cohomology is \cite{loday}[prop. 1.5.8]
\begin{displaymath}
HH^*(A) \: = \: {\rm Ext}^*_{A \otimes A^{op}}\left( A, A \right)
\end{displaymath}

In the case of the B model, the open string algebra is ${\cal O}_X$,
as the B-branes are coherent ${\cal O}_X$-sheaves. 
Since the local Ext groups reduce to Hochschild cohomology for algebras,
one defines the Hochschild cohomology of $X$ to be
\cite{andrei-utah}[def'n 6.2]
\begin{displaymath}
{\rm Ext}^*_{X\times X}\left( \Delta, \Delta 
\right)
\end{displaymath}
where, as usual, $\Delta$ denotes $\Delta_* {\cal O}_X$.
A natural guess would be that the Hochschild homology should be defined
similarly as a global Tor group; however, global Tor groups for sheaves do
not seem to be consistently defined.  It is true, however, true that
\begin{displaymath}
{\rm Ext}^*_X\left( A^{\vee}, B \right) \: = \:
{\bf R} \Gamma\left( X, A \otimes B \right)
\end{displaymath}
for any two elements $A, B \in D^b(X)$, and the right-hand-side of that
expression is morally, if not literally, a global Tor group.
Thus, the reader should not be surprised to learn that
Hochschild homology of $X$ is defined to be
\cite{andrei-utah}[def'n 6.2]
\begin{displaymath}
HH_*(X) \: = \: {\rm Ext}^{-*}_{X \times X}\left( \Delta^{\vee}, 
\Delta \right)
\end{displaymath}
(The change in sign of the grading is the convention in this context.
Although Hochschild cohomology of a space is always in positive degrees,
Hochschild homology is in both positive and negative degrees.)

In section~\ref{bmodel-closed} we saw from folding operations that
the closed string states in the B model are given by
Hochschild homology,\begin{displaymath}
HH_*(X) \: = \:
{\rm Ext}^{-*}_{X \times X} \left( \Delta^{\vee},
\Delta \right)
\end{displaymath}
ultimately because a closed string can be flattened into
an open string on $X \times X$ with diagonal boundary conditions.
Let us compare this to more standard expressions for states in the
closed string B model.

The Hochschild homology and cohomology can be related to differential
forms \cite{andrei-utah}[section 6.4], 
thanks to the ``Hochschild-Kostant-Rosenberg isomorphism''
\cite{hkr-orig,swan1,kont-hkr,yeku1}, which says that for $X$ a smooth
quasi-projective variety over ${\bf C}$,
\begin{eqnarray}
HH^n(X) \: \equiv \:
{\rm Ext}^n_{X \times X}\left( \Delta, \Delta \right)
& \cong & \bigoplus_{p+q=n} H^p\left( X, \Lambda^q TX \right)
\label{hkr1} \\
HH_n(X) \: \equiv \:
{\rm Ext}^{-n}_{X \times X}\left( \Delta^{\vee}, \Delta \right)
& \cong & \bigoplus_{q-p=n} H^p\left( X, \Omega^q_X \right)
\label{hkr2}
\end{eqnarray}
Using the fact that
\begin{displaymath}
\Omega^q_X \: = \: K_X \otimes \Lambda^{n-q} TX
\end{displaymath}
where $n$ is the dimension of $X$, we see that
\begin{displaymath}
H^p\left( X, \Omega^q_X \right) \: = \: 
H^p\left( X, K_X \otimes \Lambda^{n-q} TX \right)
\end{displaymath}
and so
\begin{eqnarray*}
HH_*(X) & = &
\bigoplus_{q-p=*} H^p\left( X, \Omega^q_X \right) \: = \:
\bigoplus_{q-p=*} H^p\left( X, K_X \otimes \Lambda^{n-q} TX \right) \\
& = & \bigoplus_{p+q=n-*} H^p\left( X, K_X \otimes \Lambda^q TX \right)
\end{eqnarray*}
When $X$ is Calabi-Yau, we see from the above that
\begin{displaymath}
HH_*(X) \: = \: HH^{n-*}(X)
\end{displaymath}
If $K_X$ is nontrivial but 2-torsion, then the Hochschild homology and
cohomology no longer have a simple relationship.

\subsection{String topology}
\label{sec:string-topology}

In \S\ref{bmodel:string-states} we used taffy identities built from
defects to explain why the closed string in a topological field theory
should be the Hochschild homology of the open string states. 

We discussed an analogue of the Hochschild-Kostant-Rosenberg isomorphism
for string topology in section~\ref{string-top}.
Specifically, we argued there that
\begin{displaymath}
HH_*(C^*X) \: = \: H^*(LX) \: = \: 
{\bf R}{\rm Hom}_X\left( C^* \Delta^{\vee}, C^* \Delta \right)
\end{displaymath}
This tells us, in part, that
open string states on $X \times X$ match
closed string states on $X$, 
but more to the point, it identifies closed string states
with the Hochschild homology
(instead of Hochschild cohomology) of $C^*X$, the open string algebra.
Just as in the B model, the closed string states are naturally associated
to Hochschild homology instead of cohomology, in this framework.

\subsection{Generalized cohomology}

Our analysis of the relationship between Hochschild homology and
closed strings has the virtue that one can insert other homotopy
functors which preserve derived pull-backs, 
yielding potentially other topological field theories.  For example,
replacing singular cochains with maps to $\mathbb{Z}\times BU$, 
one can hope to build
a topological field theory based on $K$-theory, whose closed string
algebra is the Hochschild homology $HH_{*} (K^{X})$.  In fact Hopkins
announced a result like this at the Fields Institute
\cite{hopkins-lurie}.  Note that some care must be taken in
interpreting this assertion: for example the usual analysis of
Hochschild homology and loop spaces requires the space $X$ to be
simply connected; in this context, it also suggests that one might
have to use ``connective'' $K$-theory.

One situation to which this analysis of Hochschild homology does \emph{not}
apply directly is the $B$-model considered in \S\ref{bmodel:string-states}.  
We could analyze that case in this language by using the ideas outlined
in the introduction.  Specifically, consider a ``closeable'' configuration of
$n$ open strings of the form
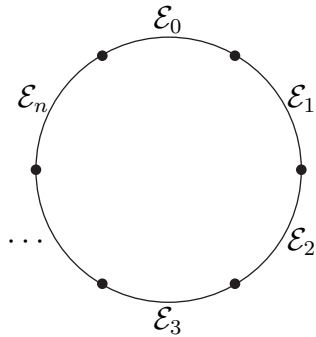
\begin{center}
\begin{picture}(120,120)
\CArc(60,60)(50,0,360)
\Vertex(110,60){2} 
\Vertex(85,103){2} 
\Vertex(35,103){2}
\Vertex(10,60){2} 
\Vertex(35,17){2}
\Vertex(85,17){2}
\Text(60,112)[b]{${\cal E}_0$}
\Text(105,87)[l]{${\cal E}_1$}
\Text(105,33)[l]{${\cal E}_2$}
\Text(60,8)[t]{${\cal E}_3$}
\Text(15,33)[r]{$\cdots$}
\Text(15,87)[r]{${\cal E}_n$}
\end{picture}
\end{center}
to which we associate
\begin{displaymath}
{\rm Hom}({\cal E}_0, {\cal E}_1) \times
{\rm Hom}({\cal E}_1, {\cal E}_2) \times
\cdots \times
{\rm Hom}({\cal E}_n, {\cal E}_0)
\end{displaymath}
(Note this is not the same thing as the massless spectrum of the diagram,
but rather is a more abstract quantity.  Also note that in this
discussion, the $\mathrm{Hom}$s should be taken in the
$\infty$-category $D_{b}^{\infty} (X)$: they are really $\RHom$s.) 
A similar configuration containing only $n-1$ open strings should be associated
a product of Hom's related to the one above via the composition
\begin{displaymath}
{\rm Hom}({\cal E}_{n-1},{\cal E}_n) \times
{\rm Hom}({\cal E}_n, {\cal E}_0) \: \longrightarrow \:
{\rm Hom}({\cal E}_{n-1}, {\cal E}_0)
\end{displaymath}
in one direction, and by setting ${\cal E}_n = {\cal E}_{n-1}$ with the
identity operator in the other direction.
This gives rise to a complex
\[
\xymatrix{
{\bigoplus_{\cE_{0},\dots ,\cE_{n}}   \hom (\cE_{0},\cE_{1})\times
\hom (\cE_{1},\cE_{2})\times \dots  \times \hom (\cE_{n},\cE_{0})}
\\
{\ldots}
\\
{\bigoplus_{\cE_{0},\cE_{1} ,\cE_{2}}    \hom (\cE_{0},\cE_{1})\times
\hom (\cE_{1},\cE_{2})\times \hom (\cE_{2},\cE_{0})}
 \ar@<-3ex>[d]  \ar[d] \ar@<+3ex>[d]\\
\\
{\bigoplus_{\cE_{0},\cE_{1}}   \hom (\cE_{0},\cE_{1}) \times \hom
(\cE_{1},\cE_{0})} 
 \ar@<-2ex>[d]  \ar@<+2ex>[d]\\
{\bigoplus_{\cE_{0}}\hom (\cE_{0},\cE_{0})}
}
\]
In this complex the structure maps are those of the Hochschild
complex, and (modulo important technical subtleties)
McCarthy and Keller \cite{McCarthy,Keller1,Keller2} use this complex to define
the Hochschild homology of the category of $\O_{X}$-modules, in such a way that
$HH_{*} (\Mod{\O_{X}})$ is the Hochschild homology of $X$.

\section{A model}
\label{A-model}

So far we have discussed the B model and string topology in this paper.
Let us now turn to the analogous constructions in the A model,
and briefly outline the highlights.
A complete analysis would involve working through the analogous
constructions in derived Fukaya categories, which we have
not done; instead, we will work solely in a large-radius limit
(hence turning off quantum corrections), and only consider a few
special cases.

Let us first consider folds in the A model.
Open strings in the A model have boundaries on either
Lagrangian or co-isotropic submanifolds, depending upon the
Chan-Paton factors.  If we let $\omega$ denote the symplectic form on
$X$, then because the folding operation reverses the orientation on the
second sheet\footnote{
We can see this by following the orientation on a string; across a fold,
the direction flips.
In the B model, this is a complex conjugation,
and this is one way of understanding why the Chan-Paton
factors on one sheet are defined by the dual bundle.
},
the symplectic form on $X \times X$ is
\begin{displaymath}
\pi_1^* \omega \: - \: \pi_2^* \omega
\end{displaymath}
With respect to this\footnote{
It is straightforward to check that without the relative sign,
{\it i.e.} for $\pi_1^* \omega + \pi_2^* \omega$,
the diagonal is not Lagrangian.
} symplectic form on $X \times X$,
the diagonal embedding $\Delta: X \rightarrow X \times X$
defines a Lagrangian\footnote{
More generally,
if $W$, $Y$ are Lagrangian in $X$, then $W \times Y$ is Lagrangian in $X     
\times X$.
} submanifold of $X \times X$ for any $X$,
and so partly as a result that submanifold together with a trivial
line bundle define a supersymmetric boundary for the A model.
That boundary will play the same role here that it did for the B model,
{\it i.e.} it will play the role of an identity in folding operations.

For the moment, let us assume all boundaries are on Lagrangian
submanifolds, for simplicity.
Omitting worldsheet instanton corrections,
the massless spectrum of an A model open string between
boundaries $(L_1, {\cal E}_1)$ and $(L_2, {\cal E}_2)$,
where $L_1$, $L_2$ are Lagrangian submanifolds of $X$,
and ${\cal E}_1$, ${\cal E}_2$ are flat vector bundles with connection
over those Lagrangian submanifolds, is given by
\begin{displaymath}
H^i_d\left( L_1 \cap L_2, {\cal E}_1^* \otimes {\cal E}_2 \right)
\end{displaymath}

Let us first consider open string states and folding tricks,
and try to re-derive a few of our results from the B model and
string topology.

Consider first a closed string on $X$ in the A model.
If we crush it to an open string on $X \times X$, as
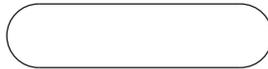
\begin{center}
\begin{picture}(100,50)
\CArc(12,24)(12,90,270)
\CArc(88,24)(12,-90,90)
\Line(12,36)(88,36)   \Line(12,12)(88,12)
\end{picture}
\end{center}
then (ignoring orientation issues on the boundaries for the moment),
we would expect that closed string states should match
\begin{displaymath}
H^*_d\left( \Delta, {\cal O} \right)
\: = \: H^*_{DR}(X)
\end{displaymath}
where $\Delta: X \hookrightarrow X \times X$ is the diagonal in $X \times X$.
Such a case has been considered previously in \cite{kont-hms}.
In fact, that reference also considered the effect of quantum
corrections (meaning, for example, that open string states are Hom's in a
derived Fukaya category), and conjectured that the product structure on
open string Hom's should match the quantum cohomology ring of $X$.

Let us also outline some of the basic manipulations of open strings,
also for simplicity in a large-radius limit where worldsheet instanton
corrections have been turned off.
Start with an oriented open string
\begin{center}
\begin{picture}(100,20)
\ArrowLine(5,10)(95,10)
\Vertex(5,10){2}  \Vertex(95,10){2}
\Text(0,10)[r]{$1$}
\Text(100,10)[l]{$2$}
\end{picture}
\end{center}
between $(L_1, {\cal E}_1)$ and $(L_2, {\cal E}_2)$, with states given by
\begin{displaymath}
H^*_d\left( L_1 \cap L_2, {\cal E}_1^* \otimes {\cal E}_2 \right)
\end{displaymath}
Now, fold this into a U-shape:
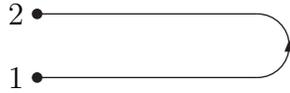
\begin{center}
\begin{picture}(100,50)
\Line(5,12)(88,12)    \Line(5,36)(88,36)
\ArrowArc(88,24)(12,-90,90)
\Vertex(5,12){2}   \Vertex(5,36){2}
\Text(0,12)[r]{$1$}
\Text(0,36)[r]{$2$}
\end{picture}
\end{center}
Following the usual procedure, open string states on the U-shaped diagram
should be given by
\begin{displaymath}
H^*_d\left( L \cap \Delta, \pi_1^* {\cal E}_1^* \otimes \pi_2^* {\cal E}_2
\right)
\end{displaymath}
where $L$ is the Lagrangian submanifold of $X \times X$ given by
\begin{displaymath}
L \: = \: \left\{ \, (x_1, x_2) \in X \times X \, | \,
x_1 \in L_1, \, x_2 \in L_2 \, \right\}
\end{displaymath}
{\it i.e.}, $L = L_1 \times L_2 \subset X \times X$.

As the physics matches, it should be true that
\begin{displaymath}
H^*_d\left( L_1 \cap L_2, {\cal E}_1^* \otimes {\cal E}_2 \right)
\: = \:
H^*_d\left( L \cap \Delta, \pi_1^* {\cal E}_1^* \otimes \pi_2^* {\cal E}_2
\right)
\end{displaymath}
and indeed, this is trivial to check.

Similarly, if we fold the same oriented string in the opposite direction
to get
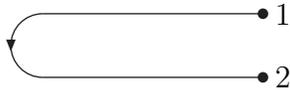
\begin{center}
\begin{picture}(100,50)
\Line(12,12)(95,12)
\Line(12,36)(95,36)
\ArrowArc(12,24)(12,90,270)
\Vertex(95,12){2}  \Vertex(95,36){2}
\Text(100,12)[l]{$2$}
\Text(100,36)[l]{$1$}
\end{picture}
\end{center}
then the corresponding open string states should again be
\begin{displaymath}
H^*_d\left( L \cap \Delta, \pi_1^* {\cal E}_1^* \otimes \pi_2^* {\cal E}_2
\right)
\end{displaymath}

So far we have merely outlined how the simplest taffy manipulations
would work in the A model, in the large-radius limit.
More generally, we conjecture that there exist analogues of the
B model taffy identities~(\ref{U-shape-main}),
(\ref{closed-U-main}), (\ref{closed-pretzel-main})
for derived Fukaya categories.

One quick check that we shall mention is to
compare to open string Gromov-Witten invariants.
We are implicitly predicting that open string Gromov-Witten invariants
of open strings on $X \times X$, with the boundary conditions determined
by the diagonal, should match closed string Gromov-Witten invariants.
(For example, a rational curve on $X$ would be a calzone-shaped object
on $X \times X$.)
Furthermore, this would also test whether the same ideas are applicable
after coupling to worldsheet gravity.  We have been informed
\cite{solomon-priv} that, indeed, open string Gromov-Witten invariants
on $X \times X$ as above
do in fact match closed string Gromov-Witten invariants on $X$,
and that there also it is merely an unravelling of definitions.

\section{Matrix factorizations}
\label{matrix-fact}

In a Landau-Ginzburg theory, we can understand a defect joining two open
strings as follows \cite{roz2,khov-roz}.
If one space is $X$ with superpotential $W_X$, and the other space is
$Y$ with superpotential $Y$, then denoting the projections from $X \times Y$
to $X$, $Y$
by $p_X$, $p_Y$, respectively, the defect is defined by a matrix factorization
in the superpotential
\begin{displaymath}
p_X^* W_X \: - \: p_Y^* W_Y
\end{displaymath}
over $X \times Y$.

We can understand the identity defect as follows.
Over the image of the diagonal embedding $\Delta: X \hookrightarrow X \times X$,
the superpotential
\begin{displaymath}
W \: \equiv \: p_1^* W_X \: - \: p_2^* W_X
\end{displaymath}
vanishes.  A matrix factorization is defined over a submanifold $S$
by a pair of vector bundles ${\cal E}$, ${\cal F}$ over $S$ with
maps $f: {\cal E} \rightarrow {\cal F}$, $g: {\cal F} \rightarrow {\cal E}$
such that $f \circ g = (W|_S) \, {\rm Id}_{\cal E}$,
$g \circ f = (W|_S) \, {\rm Id}_{\cal F}$.
In the present case, for $S$ the diagonal submanifold, $W|_S = 0$,
so we can take ${\cal E} = {\cal O}_X$, ${\cal F} = 0$,
and $f = g = 0$.

Given the structure above, we make the following conjectures for
matrix factorizations:
\begin{enumerate}
\item The hypercohomology groups
\begin{displaymath}
{\bf H}^*\left(X, \cdots \longrightarrow \: \Lambda^2 TX \:
\stackrel{dW}{\longrightarrow} \: TX \: 
\stackrel{dW}{\longrightarrow} \: {\cal O}_X \right)
\end{displaymath}
which count closed string states in Landau-Ginzburg models
\cite{gs1}
match ${\bf R}$Hom's on $X \times X$ in the category of
matrix factorizations
\begin{displaymath}
{\bf R}{\rm Hom}_{X \times X, p_1^* W - p_2^* W}\left(
(\Delta_* {\cal O}_X, 0), (\Delta_* {\cal O}_X,0) \right)
\end{displaymath}
as would be suggested from the general considerations of
section~\ref{hoch-closed}.
This has been confirmed for the local case (meaning,
affine $X$, isolated critical points) in
\cite{td} in the mathematics literature and \cite{kr} in the physics
literature.  The more general form above has also been
conjectured by others.
\item Analogues of the taffy identities~(\ref{U-shape-main}),
(\ref{closed-U-main}), (\ref{closed-pretzel-main}) hold.
\end{enumerate}

\section{Critical strings, supercritical dimensions}
\label{critical-strings}

So far we have spoken exclusively about two-dimensional
topological field theories,
but identical ideas apply, with some caveats, to full string theories.
After all, the folding operation is trivial, it is an artifact of the
worldsheet description, and does not itself convey any physics.
(The catch, the caveat, is the coupling to worldsheet gravity.
We do not understand how to couple theories with defects to
worldsheet gravity.  In this section, we will consider physical
untwisted strings, but not coupled to worldsheet gravity.)

This does lead to some counterintuitive results, however.
For example, consider a closed string on a 10-manifold $X$.
If we fold the closed string to an open string on $X \times X$,
then we have discovered that, for certain special boundary conditions,
open strings on 20-dimensional spaces behave like critical strings.

If we track through the physics in detail, we find that this is
largely correct (albeit with subtleties involving coupling to gravity).
Consider an ordinary closed string bosonic field $\phi^{\mu}$,
canonically quantized as
\begin{displaymath}
\hat{\phi}(\tau,\sigma) \: = \:
x \: + \:  \frac{p}{4 \pi} \: + \: \frac{i}{\sqrt{4\pi}} \sum_{n \neq 0}
\frac{1}{n} \left( \alpha_n e^{-in(\tau - \sigma) } \: + \:
\tilde{\alpha}_n e^{-in(\tau + \sigma)}
\right)
\end{displaymath}
Suppose these describe a local description of some manifold $X$ of dimension
$k$.  An open string on $X \times X$ with boundary conditions in the diagonal
would be described by operators
\begin{displaymath}
\hat{\phi}^{\mu}(\tau,\sigma) \: = \:
x^{\mu} \: + \:  \frac{p^{\mu}}{4 \pi} \: + \: 
\frac{i}{\sqrt{4\pi}} \sum_{n \neq 0}
\frac{1}{n} \left( \alpha_n^{\mu} e^{-in(\tau - \sigma) } \: + \:
\tilde{\alpha}_n^{\mu} e^{-in(\tau + \sigma)}
\right)
\end{displaymath}
with $\mu \in \{1, \cdots, 2k\}$, and the boundary conditions
\begin{displaymath}
\hat{\phi}^{\mu>k}(\sigma = 0,\pi) \: = \:
\hat{\phi}^{\mu-k}(\sigma=0,\pi)
\end{displaymath}
force
\begin{displaymath}
x^{\mu>k} \: = \: x^{\mu-k}, \:
p^{\mu>k} \: = \: p^{\mu-k}, \:
\alpha_n^{\mu>k} \: = \: \alpha_n^{\mu-k}, \:
\tilde{\alpha}_n^{\mu>k} \: = \: \tilde{\alpha}_n^{\mu-k}
\end{displaymath}
With these constraints, however, $\hat{\phi}$ is equivalent to a closed
string field on $X$.

So far we have demonstrated that in canonical quantization on the worldsheet,
operators on an open string on $X \times X$ with diagonal boundary
conditions are equivalent to operators on closed strings on $X$.
It is tempting to therefore conclude that physical open strings on $X \times X$
as above are therefore equivalent to physical closed strings on $X$;
however, the result should be interpreted with a certain amount of
care.
For example, the computation of the bulk central charge in the open
string theory is unaffected by the boundary conditions, and so will be
twice its value for the corresponding closed string theory.
In the present case, that might possibly signal\footnote{
We would like to thank S.~Hellerman for suggesting this possibility.
} that the number
of ghost fields on the open string worldsheet should also be doubled,
also with diagonal boundary conditions.

To properly understand the issues above would involve understanding how
defects are coupled to worldsheet gravity, which we shall not attempt here.

In passing, we feel we should mention the tangentially-related fact
that there exist tachyon-free closed string theories in dimensions
$8k+2$ \cite{hs}[section 5].

\section{Conjectures on elliptic genera}
\label{ell-gen}

Let us now apply some of these ideas to try to gain some insight
into elliptic genera.  Physically, an elliptic genus is the
torus
partition function of a (half-twisted) closed string on a space $X$.
Now, insert two identity defects along the sides, along 
which the torus can be cut open like a bagel, and glue together the
two sides to get an annulus diagram on $X \times X$.  

Now, from the Cardy condition, 
an annulus diagram can be thought of
as either a closed string propagating between two boundary states,
or an open string propagating in a loop, and furthermore that
either method of computation should give the same result.

For example, for an ordinary annulus diagram in the B model, the Cardy condition
reduces to Hirzebruch-Riemann-Roch \cite{aw}.
In the present case, we have a half-twisted closed string one-loop
diagram on $X$, represented as an annulus diagram on $X \times X$.
Following the same philosophy of annulus diagrams, it is natural to
try to
interpret the partition function of the annulus as describing an
index computation on $X \times X$.

For the right notion of open string states,
index theory should recover elliptic genera.  For example, following
\cite{ed-ellgen}, a prototype for the elliptic genus is the
expression
\begin{displaymath}
\sum q^{k/2} R_k \: = \: \bigotimes_{k=1/2,3/2,5/2,\cdots} \Lambda_{q^k} TX
\bigotimes_{\ell=1,2,3,\cdots} S_{q^{\ell}} TX
\end{displaymath}
where
\begin{eqnarray*}
R_0 & = & 1 \\
R_1 & = & TX \\ 
R_2 & = & \Lambda^2 TX \oplus TX \\
R_3 & = & \Lambda^3 TX \oplus \left( TX \otimes TX \right) \oplus
TX
\end{eqnarray*}
and so forth.  The $R_i$ are defined by the string states; the expression
above defines an index of the form
\begin{displaymath}
\sum_{i,k} q^{k/2} (-)^i {\rm dim} \: H^i(X, R_i) \: = \:
\int_X {\rm Td}(TX) \wedge {\rm ch}\left(
\bigotimes_{k=1/2,3/2,5/2,\cdots} \Lambda_{q^k} TX
\bigotimes_{\ell=1,2,3,\cdots} S_{q^{\ell}} TX
\right)
\end{displaymath}
The right-hand side we could interpret as closed strings on $X \times X$
propagating between two copies of the diagonal; the left-hand side we could
interpret as the partition function of open string states.

So far, however, this picture of elliptic genera is not particularly
noteworthy.  We can try to significantly generalize this picture as follows.
Instead of starting with an ordinary closed string,
begin with $n$ open strings on $X$
joined together along defects, defined by pushforwards along the diagonal
of $n$ elements of K theory of $X$ (and so sitting on the diagonal
embedding in $X \times X$).

We can understand those $n$ elements of K theory as
giving a finite approximation to K theory on the loop space,
in the spirit of \cite{gjp}.
(Given that elliptic genera are understood \cite{ed-ellgen} in terms of
index theory on loop spaces, this is already pertinent.)

Now, imagine propagating those defects around in a circle, to form
a torus with slanted sides.
We can use the folding tricks discussed earlier in this paper to
collapse such a diagram down to an annulus, with collections of K theory
elements on the boundaries.
If one could compute those annulus amplitudes explicitly, and furthermore
`trace over' the boundary K theory elements, then one would be able
to construct an explicit map to ordinary elliptic genera, giving an
new way to understand the relationship between elliptic genera and
K theory on loop spaces.

Although we do not have anything definitive to say here,
we feel we should observe that the work
\cite{schimm} may be relevant to such questions.

\section{Conclusions}

In this paper we have worked out mathematical identities to verify
that B model and string topology are
invariant under taffy-like operations, which involve folding and
twisting worldsheets into different-appearing yet physically-equivalent
forms.  Those identities are essentially consequences of homotopy
invariance of the underlying theory.  We have also outlined analogous
results and conjectures in other contexts, including the A model,
B-twisted Landau-Ginzburg models, and physical strings, and presented
some conjectured applications to the study of elliptic genera.

One natural speculation for future work concerns other applications
of trivial defects.  For example, although one typically only thinks of
using differential forms to compute (real-valued) cohomology on
manifolds, differential forms can also be applied to more
general topological spaces, see for example \cite{gm}.  
It is natural to speculate
whether ideas similar to those of \cite{gm} might be applied physically to 
understand string states in higher-order defect junctions.

Another direction for future work is to apply the same ideas to
topological field theories in higher dimensions, where defects have
also appeared.  Are there analogous taffy-like constructions there?

Yet another directions is to understand whether the taffy identities
presented here 
are a consequence of general mathematical axiomatic frameworks for
(enriched) topological field theories.

In a different direction, U-shaped branes have appeared in discussions
of holographic duals to chiral symmetry breaking, see {\it e.g.}
\cite{n1} and references therein.  We do not see a direct connection
 -- our work is concerned with worldsheet reparametrizations and alternate
descriptions, whereas there a brane is physically being bent -- but it
would certainly be interesting if a direct link could be found.

\section{Acknowledgements}

We would like to thank A.~Caldararu, S.~Hellerman, L.~Kong, T.~Pantev,
and I.~Runkel for
useful conversations.  In particular, A.~Caldararu checked numerous
bits of homological algebra, and S.~Hellerman made an observation about
global folding of string worldsheets that led to the present paper.
M.A. and E.S. were partially supported by NSF grant
DMS-0705381.  In addition, E.S. was partially supported by
NSF grant PHY-0755614.


\begin{thebibliography}{199}


\addcontentsline{toc}{section}{References}


\bibitem{osh-affleck} M. Oshikawa, I. Affleck, ``Boundary conformal field
theory approach to the critical two-dimensional Ising model with a
defect line,'' {\tt arXiv:  cond-mat/9612187}.

\bibitem{bachasetal} C. Bachas, J. de Boer, R. Dijkgraaf, H. Ooguri,
``Permeable conformal walls and holography,''
{\tt arXiv:  hep-th/0111210}.

\bibitem{roz1}  L. Rozansky, ``Topological A models on
seamed Riemann surfaces,'' {\tt arXiv:  hep-th/0305205}.

\bibitem{roz2} L. Rozansky, ``Topological Landau-Ginzburg models on
a worldsheet foam,''
Adv. Theor. Math. Phys. {\bf 11} (2007) 233-260,
{\tt  arXiv:  hep-th/0404189}.

\bibitem{khov-roz} M. Khovanov, L. Rozansky, ``Matrix factorizations and
link homology,'' {\tt arXiv:  math/0401268}.

\bibitem{br} I. Brunner, D. Roggenkamp, ``B-type defects in
Landau-Ginzburg models,'' {\tt arXiv:  0707.0922}.

\bibitem{fiol} B. Fiol, ``Defect CFT's and holographic multiverse,''
{\tt arXiv:  1004.0618}.

\bibitem{ksv} A. Kapustin, K. Setter, K. Vyas, ``Surface operators in
four-dimensional topological gauge theory and Langlands duality,''
{\tt arXiv:  1002.0385}.

\bibitem{bdh} A. Bartels, C. Douglas, A. Henriques,
``Conformal nets and local field theory,''
{\tt arXiv:  0912.5307}.

\bibitem{fss} J. Fuchs, C. Schweigert, C. Stigner,
``The classifying algebra for defects,''
{\tt arXiv:  1007.0401}.

\bibitem{dkr} A. Davydov, L. Kong, I. Runkel, ``Invertible defects and
isomorphisms of rational CFTs,''
{\tt arXiv:  1004.4725}.

\bibitem{s1} G. Sarkissian, ``Defects in $G/H$ coset, $G/G$ topological field
theory and discrete Fourier-Mukai transform,''
{\tt arXiv:  1006.5317}.

\bibitem{cr} N. Carqueville, I. Runkel, ``Rigidity and defect actions in
Landau-Ginzburg models,'' {\tt arXiv:  1006.5609}.

\bibitem{kaps} A. Kapustin, N. Saulina, ``Topological boundary conditions
in abelian Chern-Simons theory,'' {\tt arXiv:  1008.0654}.

\bibitem{kset} A. Kapustin, K. Setter, ``Geometry of topological defects
of two-dimensional sigma models,'' {\tt arXiv:  1009.5999}.

\bibitem{lurietft} J. Lurie, ``On the classification of
topological field theories,''
{\tt arXiv:  0905.0465}.

\bibitem{McCarthy} R. McCarthy, ``The cyclic homology of an exact category,''
J. Pure. Appl. Alg. {\bf 93} (1994) 251-296.

\bibitem{Keller1} B. Keller, ``On the cyclic homology of ringed spaces and
schemes,'' Documenta Mathematica {\bf 3} (1998) 231-259.

\bibitem{Keller2} B. Keller, ``On the cyclic homology of exact categories,''
J. Pure. Appl. Alg. {\bf 136} (1999) 1-56.

\bibitem{costello} K. Costello, ``Topological conformal field theories
and Calabi-Yau categories,'' {\tt arXiv:  math/0412149}.

\bibitem{andrei-utah} A. Caldararu, ``Derived categories of sheaves:
a skimming,'' {\tt arXiv:  math.AG/0501094}. 

\bibitem{ng} N. Ganter, M. Kapranov, ``Representation and character theory
in 2-categories,'' {\tt arXiv:  math/0602510}.

\bibitem{aw} A. Caldararu, S. Willerton, ``The Mukai pairing I:
a categorical approach,'' {\tt arXiv:  0707.2052}.

\bibitem{hopkins-lurie} M. Hopkins, J. Lurie, unpublished work, described
in a 2005 talk of M.~Hopkins at the Fields Institute

\bibitem{lurie-dag-i} J. Lurie, ``Stable infininty categories'',
{\tt arXiv: math/0608228}

\bibitem{hart-ag} R. Hartshorne, {\it Algebraic geometry}, Graduate texts
in math. 52, Springer, New York, 1977.

\bibitem{freed-ed} D. Freed, E. Witten, ``Anomalies in string theory with
D-branes,'' {\tt arXiv:  hep-th/9907189}.

\bibitem{ks} S. Katz, E. Sharpe, ``D-branes, open string vertex operators,
and Ext groups,'' Adv. Theor. Math. Phys. {\bf 6} (2003) 979-1030,
{\tt arXiv:  hep-th/0208104}.

\bibitem{serre} J. P. Serre, {\it Local algebra}, Springer monographs in
mathematics, Springer, Berlin, 2000.

\bibitem{ed-str} E. Witten, talk at Strings 2010.

\bibitem{andreipriv0} A. Caldararu, private communication.

\bibitem{ed-tft} E. Witten, ``Mirror manifolds and topological field
theory,'' {\tt arXiv:  hep-th/9112056}.

\bibitem{cmr} S. Cordes, G. Moore, S. Ramgoolam,
``Lectures on 2d Yang-Mills theory, equivariant cohomology, and 
topological field theories,'' {\tt arXiv:  hep-th/9411210}. 

\bibitem{goodwillie} T. Goodwillie, ``Cyclic homology, derivations and the
free loop space,'' Topology {\bf 24} (1985) 187-215.

\bibitem{sullivan1} D. Sullivan, ``Open and closed string field theory
interpreted in classical algebraic topology,''
{\tt arXiv:  math/0302332}.

\bibitem{sullivan2} D. Sullivan, ``String topology background and present
state,'' {\tt arXiv:  0710.4141}.

\bibitem{Adams:Cobar} J. F. Adams, ``On the cobar construction,''
Proc. Nat. Acad. Sci. USA {\bf 42} (1956), 409-412.

\bibitem{BousfieldKan} A. Bousfield and D. Kan, {\it Homotopy limits,
completions, and localiations}.  Springer Lecture Notes in Mathematics
304, 1972.

\bibitem{Rector:EM} D. Rector, ``Steenrod operations in the
Eilenberg--Moore spectral sequence'', Comm. Math. Helv. {\bf 45}
(1970), 540-552.

\bibitem{cart-eil} H. Cartan, S. Eilenberg, {\it Homological algebra},
Princeton University Press, 1956.

\bibitem{Cohen:CyclicLectureNotes} R. Cohen, K. Hess, A. Voronov,
{\it String topology and cyclic homology}, Advanced courses in
mathematics CRM Barcelona, Birkh\"auser, Basel, 2006.

\bibitem{jones} J. D. S. Jones, ``Cyclic homology and equivariant homology,''
Invent. Math. {\bf 87} (1987) 403-423.

\bibitem{gjp} E. Getzler, J. Jones, S. Petrack,
``Differential forms on loop spaces and the cyclic bar complex,''
Topology {\bf 30} (1991) 339-371.

\bibitem{bct-octft} A. J. Blumberg, R. L. Cohen, and C. Teleman,
``Open-closed field theories, string topology, and Hochschild homology,''
{\tt arXiv: 0906.5198}.

\bibitem{kr} A. Kapustin, L. Rozansky, ``On the relation between open and
closed topological strings,'' {\tt arXiv:  hep-th/0405232}.

\bibitem{ms} G. Moore, G. Segal, ``D-branes and K-theory in
2d topological field theory,'' {\tt arXiv:  hep-th/0609042}.

\bibitem{moeller-sachs} N. Moeller, I. Sachs, ``Closed string cohomology
in open string field theory,'' {\tt arXiv:  1010.4125}.

\bibitem{loday} J. L. Loday, {\it Cyclic homology}, second edition,
Springer, Berlin, 1992, 1998.

\bibitem{hkr-orig} G. Hochschild, B. Kostant, A. Rosenberg,
``Differential forms on regular affine algebras,''
Trans. AMS {\bf 102} (1962) 383-408.

\bibitem{swan1} R. G. Swan, ``Hochschild cohomology of quasiprojective
schemes,'' J. Pure Appl. Algebra {\bf 110} (1996) 57-80.

\bibitem{kont-hkr} M. Kontsevich, ``Deformation quantization of
Poisson manifolds, I,'' {\tt arXiv:  q-alg/9709040}.

\bibitem{yeku1} A. Yekutieli, ``The continuous Hochschild cochain complex
of a scheme,'' Canad. J. Math. {\bf 54} (2002) 1319-1337.

\bibitem{kont-hms} M. Kontsevich, ``Homological algebra of mirror
symmetry,'' {\tt arXiv:  alg-geom/9411018}.

\bibitem{solomon-priv} J. Solomon, private communication.

\bibitem{gs1} J. Guffin, E. Sharpe, ``A-twisted Landau-Ginzburg models,''
J. Geom. Phys. {\bf 59} (2009) 1547-1580,
{\tt arXiv:  0801.3836}.

\bibitem{td} T. Dychkerhoff, ``Compact generators in categories of
matrix factorizations,'' {\tt arXiv:  0904.4713}.

\bibitem{hs} S. Hellerman, I. Swanson, ``Charting the landscape of
supercritical string theory,'' Phys. Rev. Lett. {\bf 99} (2007) 171601,
{\tt arXiv:  0705.0980}.


\bibitem{ed-ellgen} E. Witten, ``Elliptic genera and quantum field theory,''
Comm. Math. Phys. {\bf 109} (1987) 525-536.

\bibitem{schimm} R. Schimmrigk, ``Emergent spacetime from modular motives,''
{\tt arXiv:  0812.4450}.

\bibitem{gm} P. Griffiths, J. Harris, {\it Rational homotopy theory and
differential forms}, Birkh\"auser, Boston, 1981.

\bibitem{n1} V. Niarchos, ``Hairpin-branes and tachyon-paperclips in
holographic backgrounds,'' {\tt arXiv:  1005.1650}.











\end{thebibliography}
\end{document}